\documentclass[11pt]{article}

\usepackage{amsmath,amsthm,latexsym,amssymb,amsfonts,epsfig}

\numberwithin{equation}{section}

\usepackage{graphicx} 
\usepackage{physics}
\usepackage{amssymb}
\usepackage{color}

\usepackage[hidelinks]{hyperref}
\usepackage{booktabs}

\usepackage{bm}

\allowdisplaybreaks[1]

\title{{\sf Quantum Field Theory of}\\
{\sf Black Hole Perturbations with Backreaction}\\
{\sf IV. Spherically symmetric 2nd order Einstein-Maxwell sector in generalised gauges}}

\author{
{\sf J. Neuser}$^1$\thanks{{\sf jonas.neuser@fau.de}}\\
\\
{\sf $^1$ Institute for Quantum Gravity, FAU Erlangen -- N\"urnberg,}\\
{\sf Staudtstr. 7, 91058 Erlangen, Germany}\\
}
\date{{\small\sf \today}}

\begin{document}

\maketitle

\begin{abstract}
    In previous papers of this series we analysed the reduced phase space approach to
    perturbations of Einstein-Maxwell theory to second order around spherically symmetric 
    backgrounds in the Gullstrand Painlevé Gauge and confirmed consistency with previous
    approaches. In this paper we generalize this result and show that the analysis can be 
    performed in gauges for the background variables compatible with the Gullstrand Painlevé gauge. 
    We obtain the same structure for the reduced Hamiltonian that contains 
    the well known Regge-Wheeler and Zerilli potentials. Possible applications of this 
    generalization are discussed.
\end{abstract}

\section{Introduction}

In the previous papers of this series \cite{II,III} we investigated the reduced phase space perturbations 
around spherically symmetric spacetimes in the Gullstrand-Painlev\'e gauge to second order
for Einstein-Maxwell theory. We found agreement of our method, capable to deal with
gauge invariant perturbations to any finite order \cite{I}, with different earlier approaches that are 
applicable to only second order perturbations.    
In \cite{II,III} we constructed the reduced Hamiltonian which consists of a sum of 
several, decoupled, free scalar field Hamiltonians with  different potentials. As it turns out the potentials 
correspond to the ones obtained in the linear perturbation theory for the 
equations of motion \cite{Chandrasekhar1983}. 

In the case of the Schwarzschild gauge the Hamiltonian theory of pure gravity 
and the coupling to the electromagnetic field was first studied by Moncrief 
\cite{6,Moncrief1974a,Moncrief1974RN,Moncrief1975}. 
Later Brizuela and Martín-García discussed the general situation in \cite{7,8}, 
where the authors consider general relativity coupled to a free scalar field. 
However, for the even parity sector their results are rather complicated and their final Hamiltonian 
is not very tractable. So far, in the previous papers of the series \cite{II,III} we discussed gravity 
coupled to the electromagnetic field in the Gullstrand-Painlevé gauge
perturbed to second order in reduced phase space language. 
In \cite{I} an additional scalar field is included and the general 
framework for arbitrary orders is outlined. In this manuscript we explore the theory for generalised gauge fixings for the background.
We restrict the class of admissible gauges to gauges compatible with the Gullstrand-Painlevé gauge in the sense that the fall-off behaviour at infinity is the same. Hence, we extend the work by Moncrief to more general spherically 
symmetric backgrounds. Furthermore, we extend the results by Brizuela and Martín-García 
by considering the electromagnetic field. With the help of additional canonical transformations 
we simplify the reduced Hamiltonian as compared to
\cite{7,8}.

The reduced Hamiltonian, also in a generalised gauge, to second order can be cast into a sum of decoupled, 
free scalar fields with potentials depending on the choice of background degrees of freedom. 
In this general situation, the potentials agree with the once obtained in \cite{II,III} if we fix the Gullstrand-Painlevé gauge for the background degrees of freedom.
This confirms that the reduced phase space approach to perturbation theory also works for any possible 
choice of gauge. 
We complete the analysis by verifying that the equations of motion for our Hamiltonian match the 
results of linear perturbations of the Langrangian equations of motion quoted in the literature.

This manuscript is organized as follows: In Section \ref{sec:SSBGG}, 
we analyze the spherically symmetric background degrees of freedom and recover the 
Reissner-Nordstr{\o}m solution for black holes. After that, in Section \ref{sec:TPEMS} we 
perturb the constraints to second order in the non-symmetric degrees of freedom. 
The constraints are then solved in Section \ref{sec:ATC} and we derive a physical Hamiltonian. 
We then compare our results to the literature in Section \ref{sec:Comparison}. 
Finally, we conclude in Section \ref{sec:Conclusion}.

\section{The Spherical Symmetric Background in Generalised Gauges}
\label{sec:SSBGG}

As in the previous papers of this series the starting point for the analysis is the ADM formulation of 
General Relativity coupled to an electromagnetic field. We will use the same conventions as in the previous papers. 
The explicit details for our notation are explained in \cite{II}. Here, we will just give a short reminder. 
The phase space is given by the induced metric $m_{\mu \nu}$ and its conjugate momentum $W^{\mu \nu}$. The electromagnetic field is described by the vector potential $A_\mu$ and the conjugate electric field $E^\mu$. The Hamiltonian consists of a linear combination of three constraints. The Hamiltonian constraint $V_0$, the diffeomorphism constraint $V_\mu$ and the Gau{\ss} constraint $V_G$. 

In this section we specialise to spherical symmetric degrees of freedom. For the gravitational variables we have
\begin{equation}
    m_{33} = e^{2\mu}, \quad \quad m_{3A} = 0, \quad \quad m_{AB} = e^{2\lambda} \Omega_{AB}\,,
\end{equation}
\begin{equation}
    W^{33} = \sqrt{\Omega} \frac{\pi_\mu}{2}e^{-2\mu}, \quad \quad W^{3A} = 0, \quad \quad W^{AB} = \sqrt{\Omega} \frac{\pi_\lambda}{4}e^{-2\lambda} \Omega^{AB}\,.
    \label{eq:DefSymVars}
\end{equation}
The indices $A,B=1,2$ run over the angular components and $3$ is the radial component. The metric on the sphere is denoted by $\Omega_{AB}$ and $\sqrt{\Omega}$ is an abbreviation for $\sqrt{\det(\Omega)}$. 
In the electromagnetic sector we only have the radial component of the electric field $E^3 = \sqrt{\Omega} \xi$, where $\xi$ is a constant proportional to the electric charge. 

Before we specialised to the Gullstrand-Painlevé gauge ($\mu=0$ and $\lambda=\log(r)$). We now leave the functions $\mu$ and $\lambda$ arbitrary. 
However, in order to fully study the Hamiltonian theory including boundary terms we still have to specify fall-off conditions of the canonical variables. 
For consistency with the previous papers we choose the fall-off conditions to be compatible with the Gullstrand-Painlevé gauge.
This allows us to compare the results of this paper with the results of the previous papers \cite{II,III}. We choose
\begin{align}
    \mu &\sim \mu_\infty r^{-2},\quad \quad \quad \quad \quad \quad \quad ~~\pi_\mu \sim \sqrt{\Omega}\pi_\mu^{\infty} \sqrt{r}\\
    \lambda &\sim \log(r) + \lambda_\infty r^{-2},\quad \quad \quad \quad 
    \pi_\lambda \sim \sqrt{\Omega} \pi_\lambda^{\infty} \sqrt{r}
    \label{eq:FallOffBackground}
\end{align}
where $\mu_\infty,\pi_\mu^\infty$ and $\lambda_\infty, \pi_\lambda^\infty$ are  constants with respect to $r$.

For a generalised gauge, the spherically symmetric diffeomorphism and Hamiltonian constraints take the form
\begin{align}
    {}^{(0)}C_v &= 4\pi \qty[e^{-\mu - 2\lambda} \qty(\frac{\pi_\mu^2}{8} - \frac{\pi_\mu \pi_\lambda}{4} + 2 e^{4\lambda}\qty(2 \lambda'' + 3 (\lambda')^2 - 2 \lambda' \mu' - e^{2(\mu - \lambda)})) + \frac{g^2}{2} e^{\mu - 2 \lambda} \xi^2]\,,\\
    {}^{(0)}C_h &= 4\pi \qty[\mu' \pi_\mu + \lambda' \pi_\lambda - \pi_\mu']\,.
\end{align}
We are still able to solve these constraints for $\pi_\mu$ and $\pi_\lambda$. The strategy for this is the same as before. First, we solve ${}^{(0)}C_h$ for $\pi_\lambda$:
\begin{equation}
    \pi_\lambda = \frac{1}{\lambda'}(\pi_\mu' - \mu' \pi_\mu)\,,
    \label{eq:SolPiLambdaBack}
\end{equation}
Then, we use this expression in the Hamiltonian constraint to derive a differential equation for $\pi_\mu$. We notice that the additional terms compared to the previous paper are proportional to a total derivative. We have
\begin{equation}
    \frac{\pi_\mu^2}{8} - \frac{\pi_\mu}{4 \lambda'}\qty(\pi_\mu' - \mu' \pi_\mu) + 2 \frac{e^{2\mu + \lambda}}{\lambda'}\dv{r}\qty((\lambda')^2 e^{-2\mu + 3 \lambda} - e^{\lambda} - \frac{g^2}{4} e^{-\lambda} \xi^2) = 0\,.
\end{equation}
The presence of the total derivative term  allows us to solve the differential equation for $\pi_\mu$ explicitly.
We multiply the equation with an appropriately chosen integrating factor and obtain a new differential equation
\begin{align}
    \dv{r}\qty[e^{-2\mu - \lambda}\pi_\mu^2 - 16 \qty((\lambda')^2 e^{-2\mu + 3\lambda} - e^{\lambda}- \frac{g^2}{4} e^{-\lambda} \xi^2)]=0\,.
\end{align}
The solution of this equation is straight forward and given by
\begin{align}
    \pi_\mu^2 = e^{2\mu + \lambda}\qty[16 r_s+ 16 \qty((\lambda')^2 e^{-2\mu + 3 \lambda} - e^{\lambda} - \frac{g^2}{4} e^{-\lambda} \xi^2)]\,,
\end{align}
where $r_s$ is an integration constant. We derive the solution for $\pi_\lambda$ from the Hamiltonian constraint ${}^{(0)} C_v$.
\begin{align}
    \pi_\lambda &= \frac{1}{2\pi_\mu}\qty(\pi_\mu^2 + 16 e^{4\lambda}\qty(2 \lambda'' + 3 (\lambda')^2 - 2 \lambda' \mu' - e^{2(\mu - \lambda)}) + 4 g^2 e^{2\mu} \xi^2)\\
    &= \frac{16e^{2\mu + \lambda}}{2\pi_\mu}\qty[ r_s + 2 e^{-2\mu + 3\lambda}\qty(\lambda'' + 2 (\lambda')^2 - \lambda' \mu' - e^{2(\mu - \lambda)})]
\end{align}

As a check for consistency we can insert the GP gauge condition ($\mu=0$ and $\lambda = \log(r)$). 
Then, the equations reduce to the ones found in the previous papers.

\section{The perturbed Einstein-Maxwell system}
\label{sec:TPEMS}
We generalize the spherically symmetric degrees of freedom to include asymmetric perturbations. The ansatz for 
these variables in terms of tensor spherical harmonics is the same as before. The only difference to the formulas 
provided in \cite{II,III} are the zeroth order contributions. We have
\begin{align} 
    m_{33} &= e^{2\mu} +  \sum_{l\geq 1, m} \bm x^v_{lm} L_{lm}\\
    m_{3A} &= 0 + \sum_{l \geq 1, m,I} \bm x^I_{lm} [L_{I,lm}]_A\\
    m_{AB} &= e^{2\lambda} \Omega_{AB} + \sum_{l\geq 1} \bm x^h_{lm} \Omega_{AB} L_{lm} + \sum_{l \geq 2, m, I} \bm X^I_{lm} [L_{I,lm}]_{AB}\\
    W^{33} &= \sqrt{\Omega}\qty(\frac{\pi_\mu}{2}e^{-2\mu} + \sum_{l\geq 1, m} \bm y_v^{lm} L_{lm})\\
    W^{3A} &= \sqrt{\Omega}\qty(0 +  \frac{1}{2}\sum_{l \geq 1, m,I}\bm y_I^{lm} L^A_{I,lm})\\
    W^{AB} &= \sqrt{\Omega}\qty(\frac{\pi_\lambda}{4}e^{-2\lambda} \Omega^{AB} L_{lm} + \frac{1}{2}\sum_{l\geq 1,m} \bm y_h^{lm} \Omega^{AB} + \sum_{l \geq 2, m, I} \bm Y_I^{lm} L^{AB}_{I,lm})\,.
\end{align}
For consistency with the previous papers of this series we use the same fall-off conditions of the canonical variables. In the limit $r$ to infinity, the perturbations $\delta m_{\mu \nu}$ and $\delta W^{\mu \nu}$ of $m_{\mu \nu}$ and $W^{\mu \nu}$ behave as
\begin{align}
    \label{eq:Decay}
    \begin{split}
    \delta m_{33} &\sim \delta m_{33}^+ r^{-1} + \delta m_{33}^- r^{-2}\\
    \delta m_{3A} &\sim \delta m_{3A}^+ + \delta m_{3A}^- r^{-1}\\
    \delta m_{AB} &\sim \delta m_{AB}^+ r + \delta m_{AB}^-\\
    \delta W^{33} &\sim \delta W^{33}_- + \delta W^{33}_+ r^{-1}\\
    \delta W^{3A} &\sim \delta W^{3A}_- r^{-1} + \delta W^{3A}_+ r^{-2}\\
    \delta W^{AB} &\sim \delta W^{AB}_- r^{-2} + \delta W^{AB}_+ r^{-3}\,.
    \end{split}
\end{align}
The notation on the right-hand side is the same as before: The $\delta m_{\mu \nu}^\pm$ and $\delta W^{\mu \nu}_\pm$ are constant with respect to $r$. The labels $\pm$ stand for even and odd parity with respect to the parity operator. This should not be confused with the notion of even/odd parity in the rest of this manuscript. For more details on this see \cite{II}. 

For the electromagnetic field the expansion in terms of scalar and vector spherical harmonics is
\begin{align}
    A_3 &= \sum_{l\geq 1, m} \bm x_M^{lm} L_{lm}\\
    A_B &= \sum_{l\geq 1, m,I} \bm X_M^{I,lm} [L_{I,lm}]_B\\
    E^3 &= \sqrt{\Omega}  \qty(\xi + \sum_{l\geq 1, m}\; \bm y^M_{lm} L_{lm})\\
    E^B &= \sqrt{\Omega} \sum_{l\geq 1, m,I}\bm Y^M_{I,lm} L_{I,lm}^B 
\end{align}
In accordance with the definitions in \cite{III}, the fall-off conditions of the electromagnetic degrees of freedom are
\begin{align}
    \label{eq:FAllOffEM}
    \begin{split}
    \delta A_3 &\sim \delta A_3^+ r^{-1} + \delta A_3^- r^{-2}\\
    \delta A_B &\sim \delta A_B^+ + \delta A_B^- r^{-1}\\
    \delta E^3 &\sim \delta E^3_- + \delta E^3_+ r^{-1}\\
    \delta E^B &\sim \delta E^B_- r^{-1} + \delta E_B^+ r^{-2}
    \end{split}
\end{align}

In the following we present the first and second order constraints in a generalised gauge. Let us first look 
at the first order Gauß constraint. It is given by the same expression as in \cite{III}:
\begin{equation}
    ({}^{(1)}V_G)_{lm} = \sqrt{\Omega}((\bm y^M_{lm})' - \sqrt{l(l+1)}\bm Y^M_{e,lm})
\end{equation}
We use the same solution of this constraint for $\bm y^M_{lm}$ as before. We have
\begin{equation}
    \bm y^M_{lm} = \sqrt{l(l+1)} \int \bm Y^M_{e,lm}\dd{r}\,.
\end{equation}
For the conjugate variable $\bm x^M_{lm}$ we choose the gauge fixing $\bm x^M_{lm}=0$.

After the successful treatment of the Gauß constraint, we consider the Hamiltonian and diffeomorphism constraints next. We suppress the labels $l$ and $m$ in the following equations. 
The convention for dropping the indices is the same as in the previous papers. 
The derivation of the perturbed constraints to second order is possible by hand. Some intermediate steps of the computation are documented in appendix \ref{sec:Expansion2ndOrder}. The result was checked with Mathematica and the xAct package \cite{xAct}. The first order non-symmetric constraints are given by
\begin{align}
    \begin{split}
    {}^{(1)}Z^h_{lm} &= -2 e^{\mu} \partial_r (e^{\mu} \bm y_v) + e^{2\mu}\sqrt{l(l+1)} \bm y_e + 2 \lambda' e^{2 \lambda}\bm y_h - \partial_r( \pi_\mu e^{-2 \mu}) \bm x^v - \frac{1}{2} \pi_\mu e^{-2 \mu}\partial_r \bm x^v\\
    &+\frac{\pi_\lambda}{2} e^{-2\lambda} \sqrt{l(l+1)} \bm x^e + \frac{\pi_\lambda}{2} e^{-2 \lambda} \partial_r \bm x^h
    \end{split}\\
    \begin{split}
    {}^{(1)}Z^e_{lm} &= \sqrt{2(l+2)(l-1)}\qty(e^{2\lambda} \bm Y_e + \frac{\pi_\lambda}{4}e^{-2 \lambda}\bm X^e) - \partial_r \qty(e^{2\lambda} \bm y_e + \pi_\mu e^{-2 \mu}\bm x^e) - \sqrt{l(l+1)} e^{2 \lambda} \bm y_h\\
    &+ \frac{1}{2} \sqrt{l(l+1)} \pi_\mu e^{-2 \mu} \bm x^v - \xi \partial_r \bm X^e_M 
    \end{split} \\
    {}^{(1)}Z^o_{lm} &= \sqrt{2(l+2)(l-1)}\qty(e^{2\lambda} \bm Y_o +  \frac{\pi_\lambda}{4}e^{-2\lambda} \bm X^o) - \partial_r \qty(e^{2\lambda} \bm y_o + \pi_\mu e^{-2\mu}\bm x^o ) - \xi \partial_r \bm X^o_M\\
    \begin{split}
    {}^{(1)}Z^v_{lm} &=  \frac{1}{2} (\pi_\mu - \pi_\lambda) e^{\mu - 2 \lambda} \bm y_v- \frac{1}{2} \pi_\mu e^{-\mu} \bm y_h - \frac{1}{8} \pi_\mu^2 e^{-\mu - 4 \lambda} \bm x^h + \frac{1}{16}(3 \pi_\mu^2 - 2 \pi_\mu \pi_\lambda) e^{-3 \mu-2\lambda} \bm x^v\\
    &+ e^{-3\mu + 2\lambda} ( - 2 \lambda' \partial_r + 6 \lambda' \mu' - 3 (\lambda')^2 - 2 \lambda'' - e^{2\mu -2 \lambda}(l(l+1) +1)) \bm x^v\\
    &+ 2e^{-\mu}\qty(\partial_r^2 - \mu' \partial_r - \lambda' \partial_r + (\lambda')^2  - \frac{1}{2} e^{2\mu - 2\lambda}l(l+1)) \bm x^h+ 2 e^{-\mu}\sqrt{l(l+1)} (\partial_r - \mu' + \lambda') \bm x^e \\
    &-e^{\mu - 2 \lambda}\sqrt{\frac{(l+2)(l+1)l(l-1)}{2}} \bm X^e  + \frac{g^2}{2}e^{-\mu - 2 \lambda}\qty(\qty(\frac{1}{2}\bm x^v - \bm x^h e^{2(\mu -\lambda)}) \xi^2 + 2 e^{2\mu} \xi \pi_\alpha)
    \end{split}
\end{align}
For the second order symmetric constraints we obtain
\begin{align}
    \begin{split}
    \label{eq:Ch2}
    {}^{(2)}C_h &= - \bm x^o \cdot \partial_r \bm y_o + \bm Y_o \cdot \partial_r \bm X^o + \bm y_v \cdot \partial_r \bm x^v - 2 \partial_r (\bm x^v \cdot \bm y_v) - \bm x^e \partial_r \cdot \bm y_e + \bm Y_e \cdot \partial_r \bm X^e + \bm y_h \cdot \partial_r \bm x^h\\
    &+ \bm Y^M_e \cdot \partial_r \bm X^e_M + \bm Y^M_o \cdot \partial_r \bm X^o_M
    \end{split}\\
    \label{eq:Cv2}
    {}^{(2)}C_v &= \frac{e^\mu}{2} \bm y_o \cdot \bm y_o + e^{-\mu + 2 \lambda} \bm Y_o \cdot \bm Y_o  + \frac{1}{2}e^{-\mu - 2 \lambda}\pi_\mu \bm x^o \cdot \bm y_o  - \frac{1}{2}e^{-\mu - 2 \lambda}(\pi_\mu - \pi_\lambda) \bm Y_o \cdot \bm X^o\nonumber\\
    &+ \qty(\frac{\pi_\mu^2}{16} + \frac{\pi_\mu \pi_\lambda}{8}) e^{-3\mu - 4 \lambda} \bm x^o \cdot \bm x^o - e^{-3\mu}\bm x^o\cdot \qty(-4 \lambda' \partial_r - 2\lambda'' + (\lambda')^2 + 6 \lambda' \mu' -\frac{1}{2}e^{2(\mu-\lambda)}l(l+1))\bm x^o\nonumber\\
    &-e^{-\mu - 2 \lambda} \sqrt{\frac{(l+2)(l-1)}{2}}\bm x^o \cdot \qty( \partial_r -2 \lambda') \bm X^o + \qty(\frac{\pi_\mu^2}{32} - \frac{\pi_\mu \pi_\lambda}{16} + \frac{\pi_\lambda^2}{16})e^{-\mu - 6\lambda}\bm X^o \cdot \bm X^o\nonumber\\
    &-e^{-2 \lambda -  \mu} \bm X^o \cdot \qty(\partial_r^2 - 4 \lambda' \partial_r - \mu' \partial_r + \frac{5}{2} (\lambda')^2 + \lambda' \mu'  - \lambda'')\bm X^o - \frac{3}{4}e^{-2 \lambda - \mu} \partial_r \bm X^o \cdot \partial_r \bm X^o\nonumber\\
    &+\frac{1}{2} e^{3\mu - 2 \lambda}\bm y_v \cdot \bm y_v + \frac{1}{4}e^{-\mu-2\lambda}\bm x^v \cdot \bm y_v(3\pi_\mu - \pi_\lambda) + e^{-5\mu - 2 \lambda}\bm x^v \cdot \bm x^v \qty(\frac{3}{64}\pi_\mu^2 + \frac{1}{32}\pi_\mu \pi_\lambda)\nonumber\\
    &+ \frac{1}{4}e^{-5\mu + 2 \lambda}\bm x^v \cdot \qty(e^{2(\mu - \lambda)} + 12 \lambda' \partial_r + 6 \lambda'' -30 \lambda' \mu' + 9 (\lambda')^2)\bm x^v + \frac{1}{2} e^{\mu} \bm y_e \cdot \bm y_e + \frac{1}{2} \pi_\mu e^{-\mu - 2 \lambda} \bm x^e \cdot \bm y^e\nonumber\\
    &+ \qty(\frac{1}{16}\pi_\mu^2 + \frac{1}{8} \pi_\mu \pi_\lambda) e^{-3\mu - 4\lambda} \bm x^e \cdot \bm x^e + e^{-3\mu} \bm x^e \cdot \qty(4 \lambda' \partial_r + 2 \lambda'' - 6 \lambda' \mu' - (\lambda')^2)\bm x^e+\frac{1}{8}e^{-\mu - 6\lambda}\pi_\mu^2 \bm x^h \cdot \bm x^h\nonumber\\
    &- \frac{1}{2}e^{-\mu - 2 \lambda} \qty(4 (\lambda')^2\bm x^h\cdot \bm x^h - 4\lambda' \bm x^h \cdot \partial_r \bm x^h +\partial_r \bm x^h \cdot \partial_r \bm x^h)\nonumber\\
    &+e^{-\mu + 2 \lambda}\bm Y_e \cdot \bm Y_e - \frac{1}{2}e^{-\mu - 2 \lambda}  (\pi_\mu - \pi_\lambda) \bm X^e \cdot \bm Y_e + e^{-\mu - 6 \lambda}\qty(\frac{1}{32}\pi_\mu^2 - \frac{1}{16}\pi_\mu \pi_\lambda + \frac{1}{16}\pi_\lambda^2) \bm X^e\cdot \bm X^e\\
    &+ e^{\mu - 2\lambda} \frac{1}{4} \qty( - 3\partial_r \bm X^e \cdot \partial_r \bm X^e + \bm X^e\cdot(- 4\partial_r^2 + 4 \mu' \partial_r + 16 \lambda' \partial_r - 4 \mu' \lambda' - 10 (\lambda')^2+ 4 \lambda'')\bm X^e)\nonumber\\
    & - e^{\mu}\bm y_h \cdot \bm y_v - \frac{\pi_\mu}{2}e^{\mu - 4 \lambda}\bm y_v \cdot \bm x^h - \frac{1}{4}\pi_\mu e^{-3\mu} \bm x^v \cdot \bm y_h - \frac{3}{16}\pi_\mu^2 e^{-3\mu - 4 \lambda} \bm x^v \cdot\bm x^h\nonumber\\
    & - e^{- 3\mu } \bm x^v \cdot \Big(\frac{1}{2} l(l+1) e^{2(\mu - \lambda)} \bm x^h + \partial_r^2 \bm x^h - 3  \mu' \partial_r \bm x^h +(\lambda')^2 \bm x^h - \lambda' \partial_r \bm x^h)- e^{-3\mu} \partial_r \bm x^h \cdot \partial_r \bm x^v\nonumber\\
    & - \sqrt{\frac{(l+2)(l+1)l(l-1)}{8}} e^{-\mu - 2 \lambda}\bm x^v \cdot \bm X^e - e^{-3 \mu }\sqrt{l(l+1)}\bm x^v \cdot\partial_r \bm x^e\nonumber\\
    &-  e^{-3\mu}\sqrt{l(l+1)} \bm x^e \cdot\qty(- 3 \mu' \bm x^v - \lambda' \bm x^v + \partial_r \bm x^v) + e^{-\mu - 2 \lambda}\sqrt{l(l+1)} \bm x^e \cdot \qty(2 \lambda' \bm x^h - \partial_r \bm x^h)\nonumber\\
    &+ \sqrt{\frac{(l+2)(l-1)}{2}} e^{-\mu - 2 \lambda} \bm x^e \cdot \qty( 2 \lambda' \bm X^e - \partial_r \bm X^e)\nonumber\\
    & + \frac{1}{2}g^2 e^{-\mu -2 \lambda}\Bigg[\frac{1}{8} \Big(- e^{-4\mu}\bm x^v \cdot \bm x^v - 4 e^{-2\mu -2 \lambda} \bm x^v \cdot \bm x^h + 4 e^{-2\mu -2 \lambda}(\bm x^o\cdot \bm x^o+ \bm x^e \cdot \bm x^e)+ 8 e^{-4\lambda} \bm x^h \cdot \bm x^h\nonumber\\
    &~~~~~~~~~+ 2 e^{-4\lambda} (\bm X^e \cdot \bm X^e + \bm X^o \cdot \bm X^o)\Big) e^{2\mu} \xi^2 + \sqrt{l(l+1)} \xi (\bm x^v - 2 e^{2 \mu -2\lambda} \bm x^h) \cdot \int \bm Y_e^M \dd{r}\nonumber\\
    &~~~~~~~~~ + 2 \xi \bm x^o  \cdot \bm Y_o^M+ 2 \xi \bm x^e \cdot \bm Y_e^M + e^{2\mu} l(l+1) \int \bm Y_e^M \dd{r} \cdot \int \bm Y_e^M \dd{r} + e^{2\lambda} \qty( \bm Y_o^M \cdot \bm Y_o^M + \bm Y_e^M \cdot \bm Y_e^M) \Bigg] \nonumber\\
    &+ \frac{1}{2g^2}e^{-\mu - 2 \lambda} \qty[e^{2\mu}l(l+1)\bm X^o_M \cdot \bm X^o_M + e^{2\lambda} \qty(\bm X^o_M{}' \cdot \bm X^o_M{}' + \bm X^e_M{}' \cdot \bm X^e_M{}')]\nonumber
\end{align}

\section{Analysis of the Constraints}
\label{sec:ATC}

The next step of the program is the in depth analysis of the constraints. We follow the formalism outlined in \cite{TT}. 
The general strategy is the same as in the previous papers. The difference is that the canonical transformations 
need to be generalized and the necessary computations are more involved.

\subsection{The second order}
Before solving the first order constraints, we study the second order symmetric constraints. 
We work perturbatively and expand $\pi_\mu$ and $\pi_\lambda$ as $\pi_\mu = \pi_\mu^{(0)} + \pi_\mu^{(2)}$ and similarly for $\pi_\lambda$.
The quantities $\pi_\mu^{(2)}$ and $\pi_\lambda^{(2)}$ are of second order in the non-symmetric degrees of freedom. 
The expansion of the symmetric constraints to second order are
\begin{align}
    C_v &\sim \pi e^{-\mu - 2\lambda} \qty(\pi_\mu^{(0)}\pi_\mu^{(2)} - \pi_\mu^{(0)} \pi_\lambda^{(2)} - \pi_\mu^{(2)} \pi_\lambda^{(0)}) + {}^{(2)}C_v=0\,,\\
    C_h &\sim 4 \pi \qty[\mu' \pi_\mu^{(2)} + \lambda' \pi_\lambda^{(2)} - (\pi_\mu^{(2)})'] + {}^{(2)}C_h=0\,.
\end{align}
In the equation the zeroth order correction vanishes because $\pi_\mu^{(0)}$ and $\pi_\lambda^{(0)}$ are solutions of the zeroth order symmetric constraints.
$ {}^{(2)}C_v$ and ${}^{(2)}C_h$ are contributions which depend explicitly on the perturbative degrees of freedom and the zeroth order symmetric variables $\pi_\mu^{(0)}$ and $\pi_\lambda^{(0)}$.
The plan of this subsection is to solve the two equations for $\pi_\mu^{(2)}$ and $\pi_\lambda^{(2)}$. The second equation $C_h = 0$ implies
\begin{equation}
    \pi_\lambda^{(2)} = \frac{1}{\lambda'}\qty((\pi_\mu^{(2)})' - \mu' \pi_\mu^{(2)} - \frac{1}{4\pi}{}^{(2)}C_h)\,.
\end{equation}
This is used to eliminate $\pi_\lambda^{(2)}$ in the first equation $C_v=0$. The result is a differential equation for $\pi_\mu^{(2)}$:
\begin{equation}
    \frac{\pi_\mu^{(0)}}{\lambda'}\qty(\lambda' + \mu' - \lambda' \frac{\pi_\lambda^{(0)}}{\pi_\mu^{(0)}})\pi_\mu^{(2)} - \frac{\pi_\mu^{(0)}}{\lambda'}(\pi_\mu^{(2)})' + \frac{\pi_\mu^{(0)}}{4 \pi \lambda'}{}^{(2)}C_h + \frac{e^{\mu + 2 \lambda}}{\pi} {}^{(2)}C_v=0\,.
\end{equation}
The quantities $\pi_\mu^{(0)}$ and $\pi_\lambda^{(0)}$ satisfy the equations of the zeroth order. Thus, we employ equation \eqref{eq:SolPiLambdaBack} to remove $\pi_\lambda^{(0)}$ in terms of $\pi_\mu^{(0)}$ and $\pi_\mu^{(0)}{}'$. The equation can then be rewritten in the form
\begin{equation}
    (e^{-\lambda - 2 \mu}\pi_\mu^{(0)} \pi_\mu^{(2)})' - \frac{e^{-\lambda - 2 \mu}}{4\pi}\pi_\mu^{(0)}{}^{(2)}C_h - \frac{e^{- \mu + \lambda}}{\pi} \lambda' {}^{(2)}C_v=0\,.
\end{equation}
The solution of this differential equation is straight forward by integration. We have
\begin{align}
    \pi_\mu^{(2)} = \frac{e^{\lambda + 2 \mu}}{\pi \pi_\mu^{(0)}} \int \dd{r} \qty[\frac{1}{4} e^{-\lambda - 2 \mu} \pi_\mu^{(0)} {}^{(2)} C_h +  e^{-\mu + \lambda} \lambda' {}^{(2)}C_v]\,.
    \label{eq:SolIntegralpimu2}
\end{align}
From $C_v=0$ we compute $\pi_\lambda^{(2)}$ and obtain
\begin{equation}
    \pi_\lambda^{(2)} = \qty(1 - \frac{\pi_\lambda^{(0)}}{\pi_\mu^{(0)}})\pi_\mu^{(2)} + \frac{e^{\mu + 2 \lambda}}{\pi \pi_\mu^{(0)}}{}^{(2)}C_v\,.
\end{equation}
This completes the solution of the second order constraints.
It remains to determine the functions ${}^{(2)}C_v$ and ${}^{(2)} C_h$ explicitly.
In this manuscript we do not consider the dipole perturbations $l=1$ and restrict ourselves to solving the first order constraints for $l\geq2$ where both the perturbed gravitational and the perturbed electromagnetic field have true degrees of freedom. 
As in the previous papers we treat the even and odd parity sectors separately. 
In the odd parity sector we consider the variables $\bm x^o, \bm y_o$ as the gauge degrees of freedom and the variables $\bm X^o, \bm Y_o$ and $\bm X^o_M, \bm Y_o^M$ as the true degrees of freedom. 
In the even parity sector we proceed differently. We work in a different gauge where we fix $\bm X^e= \bm x^e= p_2 =0$. The quantity $p_2$ is a variable that we will introduce through a canonical transformation. In this gauge the algebraic manipulations are easier because we do not have to solve differential equations. 
In the end we comment on when it is possible to instead work with the variables $\bm X^e, \bm Y_e$ and $\bm X^e_M, \bm Y_e^M$ as the true degrees of freedom in the gauge $\bm x^h=\bm x^v=\bm x^e=0$. 

\subsection{The First Order -- Odd Parity}

In the odd parity sector there is one first order constraint, $Z^o$. The gauge degrees of freedom are $\bm x^o,\bm y_o$ and consequently we have to solve the constraint for $\bm y_o$. The solution is
\begin{align}
    \bm y_o^{(1)} = e^{-2\lambda} \int \dd{r}\qty[\sqrt{2(l+2)(l-1)}\qty(e^{2\lambda} \bm Y_o + \frac{\pi_\lambda}{4}e^{-2\lambda}\bm X^o)- \xi \partial_r \bm X^o_M]\,.
\end{align}
For the computation of $\bm y_o^{(1)}$ and in the following we work in the gauge $\bm x^o = 0$. Next, we define new variables $\tilde Q, \tilde P$ with the canonical transformation
\begin{align}
    \tilde P &:= \frac{1}{\sqrt{2}} \partial_r \qty(e^{-2\lambda} \bm X^o)\\
    \tilde Q &:= \sqrt{2} \int \dd{r}\qty( e^{2\lambda} \bm Y_o + \frac{\pi_\lambda^{(0)}}{4} e^{-2\lambda} \bm X^o)\,.
\end{align}
The motivation for this transformation comes from the integral in the solution for $\bm y_o^{(1)}$. For the electromagnetic field we introduce the quantities $A^o = \bm X^o_M$ and $\Pi_A^o = \bm Y_o^M$ to unify the notation with the even parity sector.

Next, we insert the odd parity contributions to ${}^{(2)} C_h$ and ${}^{(2)}C_v$ into the solution for $\pi_\mu^{(2)}$ in equation \eqref{eq:SolIntegralpimu2}. Then, we replace $\bm y_o$ by the result $\bm y_o^{(1)}$ and apply the canonical transformation. The resulting expression in terms of $\tilde Q, \tilde P$ and $A^o, \Pi_A^o$ is simplified using integration by parts. We end up with
\begin{align}
    &\pi_\mu^{(2)}\Big|_{\mathrm{odd}} =\frac{e^{\lambda + 2 \mu}}{\pi \pi_\mu^{(0)}} \int \dd{r} \Bigg[\frac{\pi_{\mu}^{(0)}}{4}e^{-\lambda - 2 \mu} (\tilde P \cdot \tilde Q' + \Pi_A^o \cdot (A^o)') + e^{-\mu + \lambda} \lambda'\Big(\frac{1}{2} e^{-\mu+ 2\lambda} \tilde P\cdot \tilde P\nonumber\\
    &+ \frac{e^{-\mu - 2 \lambda}}{2} \tilde Q'\cdot \tilde Q' + \frac{e^{\mu-4\lambda}}{2}(l+2)(l-1) \tilde Q \cdot \tilde Q  + \frac{g^2}{2}e^{-\mu} \Pi_A^o \cdot \Pi_A^o + \frac{1}{2g^2}e^{-\mu - 2 \lambda} \Big(e^{2\mu}l(l+1)A^o \cdot A^o\\
    &+ e^{2\mu -2 \lambda}g^2 \xi^2 A^o \cdot A^o + e^{2\lambda} {A^o}'\cdot A^o{}' - 2 \sqrt{(l+2)(l-1)} e^{2\mu -2 \lambda}  g^2  \xi \tilde Q  \cdot A^o\Big)\Big)\Bigg]\,.\nonumber
\end{align}
We did not include the boundary term from the integration by parts in the above equation. It is given by
\begin{align}
    \int \dd{r} \dv{r}\qty[2 e^{-2\mu+3\lambda} \lambda' \tilde P \cdot \int \tilde P \dd{r} + \qty(\frac{r_s}{2}- e^\lambda + 2 e^{3\lambda- 2 \mu}(\lambda')^2)\int \tilde P \dd{r} \cdot \int \tilde P \dd{r}]\,.
\end{align}
From the fall-off conditions as $r$ goes to infinity of the gravitational field we have that $\tilde P$ behaves as $r^{-2}$ and that $\tilde Q$ grows at most linearly in $r$. For the background variables we defined that $e^{\lambda}$ behaves as $r$ and that $e^\mu$ behaves asymptotically as $1 + \mu_\infty r^{-2}$ (see \eqref{eq:FallOffBackground}). Therefore, the boundary term for our choice of fall-off conditions behaves as $r^{-1}$ in the limit $r$ to infinity and can be neglected.

The solution for $\pi_\mu^{(2)}$ is further simplified using a second canonical transformation. We introduce the variables $Q^o,P^o$ for the gravitational sector. The transformation is given by a rescaling of $\tilde Q$. The variable $\tilde P$ is shifted to remove a coupling term of the form $Q^o P^o$ which would otherwise show up. We use the transformation 
\begin{align}
    \tilde Q &= e^{\lambda} Q^o\,,\\
    \tilde P &= e^{-\lambda} (P^o - \frac{\pi_\mu}{4} e^{-2\lambda} Q^o)\,.
\end{align}
We change to the new variables $Q^o,P^o$ in $\pi_\mu^{(2)}$ and simplify the expression using integration by parts. The result is
\begin{align}
    \pi_\mu^{(2)}\Big|_{\mathrm{odd}} =& \frac{e^{\lambda + 2 \mu}}{\pi \pi_\mu^{(0)}} \int \dd{r} \Bigg[\frac{1}{4}e^{-\lambda - 2 \mu} \pi_{\mu}^{(0)}(P^o \cdot (Q^{o})' + \Pi_A^o (A^o)') + \frac{1}{2} e^{-2\mu + \lambda} \lambda'\Big(P^o \cdot P^o + Q^o{}'\cdot Q^o{}'\nonumber\\
    &+ e^{2\mu-4\lambda}(l(l+1)e^{2\lambda} - 3 r_s e^{\lambda} + g^2 \xi^2) Q^{o}\cdot Q^o\Big) + \frac{1}{2}e^{-2\mu +\lambda}\lambda'\Big( g^2\Pi_A^o \cdot \Pi_A^o\\
    &+ \frac{1}{g^2}((A^o)'\cdot (A^o)' + e^{2\mu -4 \lambda}(l(l+1) e^{2\lambda} + g^2\xi^2 )A^o \cdot A^o - 2 \sqrt{(l+2)(l-1)} e^{2\mu - 3 \lambda}  \xi Q^o \cdot A^o)\Big)\Bigg]\nonumber\,.
\end{align}
The boundary term that we neglected is given by
\begin{align}
    - \frac{1}{32}\int \dd{r} \pdv{r}\qty(e^{-2\mu - 3 \lambda} (Q^o) ^2\qty(\pi_\mu^2 - 16 e^{4\lambda}(\lambda')^2))= - \int \dd{r} \pdv{r}\qty(\frac{e^{-2\lambda}}{2}\qty(r_s - e^{\lambda} - \frac{g^2 \xi^2}{4} e^{-\lambda}) (Q^o)^2)
\end{align}
From the canonical transformation we have that $Q^o$ is approaching a constant as $r$ goes to infinity. The boundary term therefore falls off like $r^{-1}$ and can be neglected in the limit.

As in the Gullstrand-Painlevé gauge, it is convenient to introduce some potentials. We have a potential for the gravitational degrees of freedom $V^o_{\mathrm{grav}}$, for the electromagnetic degrees of freedom $V^o_{\mathrm{em}}$ and for the coupling term $V^o_{\mathrm{Coup}}$. Using the potentials, the even and odd parity solutions can be treated analogously later. We define 
\begin{align}
    V_{\mathrm{grav}}^{o} &:=e^{2\mu - 2\lambda}\qty(U^o - \frac{3}{2} e^{-\lambda} r_s W^o)\\
    V_{\mathrm{em}}^{o} &:=e^{2\mu - 2\lambda}\qty(U^o + \frac{3}{2} e^{-\lambda} r_s W^o)\\
    V_{\mathrm{Coup}}^{o} &:=e^{2\mu - 3\lambda} g \xi W^o\,,
\end{align}
with the quantities
\begin{align}
    W^{o} &:= -1\\
    U^{o} &:= l(l+1) + \frac{3}{2} e^{-\lambda} r_s + e^{-2\lambda} g^2 \xi^2
\end{align}
With these potentials the solution for $\pi_\mu^{(2)}$ is 
\begin{align}
\begin{split}
    &\pi_\mu^{(2)}\Big|_{\mathrm{odd}} = \frac{e^{\lambda+2\mu}}{\pi \pi_\mu^{(0)}}\int \dd{r}\Bigg[ \frac{1}{4}e^{-\lambda - 2 \mu} \pi_{\mu}^{(0)}(P^o \cdot (Q^{o})' + \Pi_A^o \cdot (A^o)') + \frac{1}{2} e^{-2\mu + \lambda} \lambda'\Big(P^o \cdot P^o + g^2\Pi_A^o \cdot \Pi_A^o\\
    &+ Q^o{}'\cdot Q^o{}' +\frac{1}{g^2}(A^o)' \cdot (A^o)' + V_{\mathrm{grav}}^{o} Q^o \cdot Q^o + \frac{1}{g^2} V_{\mathrm{em}}^{o} A^o \cdot A^o + \frac{2}{g} \sqrt{(l+2)(l-1)} V_{\mathrm{Coup}}^{o} Q^o \cdot A^o \Big)\Bigg]\,,
    \end{split}
\end{align}
This completes the odd parity sector. Reducing all the formulas to the Gullstrand-Painlevé gauge we recover the steps in \cite{III} and are consistent with the previous papers.

\subsection{The First Order -- Even Parity}

The next objective is the analysis of the even parity sector. As explained before and similarly to the previous papers in this series we first deviate from the procedure by considering $\bm X^e,\bm Y_e$ not as true but as gauge degrees of freedom. Later we give some comments on how to go back to the original proposal where $\bm X^e, \bm Y_e$ and $\bm X^e_M, \bm Y^M_e$ are the true degrees of freedom. 

We start with the analysis of the diffeomorphism constraints ${}^{(1)}Z^h_{lm}$ and ${}^{(1)}Z^e_{lm}$. We solve them for $\bm y_e$ and $\bm Y_e$ and work in the gauge $\bm x^e=\bm X^e=0$. We obtain
\begin{align}
    \bm y_e^{(1)} =& - \frac{e^{-2\mu}}{\sqrt{l(l+1)}}\Big( -2 e^{\mu} \partial_r (e^{\mu} \bm y_v) + 2 \lambda' e^{2 \lambda}\bm y_h - \partial_r( \pi_\mu e^{-2 \mu}) \bm x^v - \frac{1}{2} \pi_\mu e^{-2 \mu}\partial_r \bm x^v + \frac{\pi_\lambda}{2} e^{-2 \lambda} \partial_r \bm x^h\Big)\\
    \bm Y_e^{(1)} =&- \frac{e^{-2\lambda}}{\sqrt{2(l+2)(l-1)}}\Big( -\partial_r(e^{2\lambda} \bm y^{(1)}_e) - \sqrt{l(l+1)} e^{2 \lambda} \bm y_h + \frac{1}{2} \sqrt{l(l+1)} \pi_\mu e^{-2 \mu} \bm x^v - \xi \partial_r \bm X^e_M\Big)
\end{align}

The remaining constraint is the first order Hamiltonian constraint ${}^{(1)}Z^v_{lm}$ which we would like to solve for $\bm x^h$. There is an obstruction to that because it depends on first and second derivatives of $\bm x^h$. To resolve this problem we introduce new variables $q_1, p_1, q_2$ and $p_2$ where we set $q_2=\bm x^h$. The transformation is tuned in such a way that the transformed constraint ${}^{(1)}Z^v_{lm}$ does not depend on the derivatives of $q_2$. We take
\begin{align}
\begin{split}
    \bm x^v &= q_1 + B q_2 + C \partial_r q_2 + D p_1\\
    \bm x^h &= q_2\\
    \bm y_v &= p_1 + G \partial_r q_2 \\
    \bm y_h &= p_2 - B p_1 + \partial_r \qty[(C - D G) p_1] - \partial_r (G q_1) + K q_2 - BG \partial_r q_2\,,
\end{split}
\end{align}
with
\begin{align}
\begin{split}
    C &= \frac{1}{\lambda'}e^{2\mu - 2 \lambda}\\
    D &= - \frac{4 \pi_\mu e^{4\mu}}{\pi_\mu^2 + 16 (\lambda')^2 e^{4\lambda}}\\
    G &= - \frac{\pi_\mu}{4\lambda'} e^{-2\mu-2\lambda}\\
    B&=\frac{8e^{4\mu}}{\pi_\mu^2 + 16 e^{4 \lambda}(\lambda')^2 }\qty(e^{-\lambda}r_s -  (l(l+1)+2))\,.
\end{split}
\end{align}
The function $K$ is longer and more complicated. We have 
\begin{align}
    K &= \frac{32 e^{-6 \lambda} e^{-2 \mu}}{M_1^2
   \pi _{\mu } (\lambda ')^4} \Bigg(2 e^{4
   \lambda-2 \mu} (\lambda ')^5 \partial_r \qty(\frac{\left(2
   \Delta^2+\Delta  \Lambda \right) e^{2 \mu}}{(\lambda ')^2}) -\Lambda  l  (l+1) \left(\Delta^2 e^{4 \mu}-3 \Delta  e^{2 \lambda+2 \mu} (\lambda')^2 + 2 e^{4 \lambda} (\lambda')^4\right)\nonumber\\
   &- 2 l(l+1) \left(2
   \Delta^2 e^{2\lambda+2 \mu} (\lambda')^2-5 \Delta  e^{4 \lambda} (\lambda')^4+4 e^{6 \lambda-2 \mu} (\lambda ')^6\right) -4 \Delta  e^{4
   \lambda} (\lambda')^4\\
   &- r_s e^{-\lambda} \left(\Delta^2 e^{2
   \lambda+2 \mu} (\lambda')^2-4 \Delta  e^{4 \lambda} (\lambda
   ')^4+2 e^{6 \lambda-2 \mu} (\lambda')^6\right)\Bigg)\nonumber\,,
\end{align} 
We introduced the quantities $\Delta, \Lambda$ and $M_1$ to simplify the formula. This notation is used throughout this section and is the generalisation of the corresponding expressions in \cite{III}. The quantities are defined as
\begin{align}
    n &:= \frac{1}{2}(l+2)(l-1)\\
    \Delta &:= 1 - e^{-\lambda}r_s + \frac{1}{4} e^{-2\lambda} g^2 \xi^2\\
    \Lambda &:= n + \frac{3}{2} e^{-\lambda} r_s - \frac{1}{2} e^{-2\lambda} g^2 \xi^2\\
    M_1 &:= \frac{4}{\lambda'} e^{-\mu - \lambda}\qty(\Delta - 2 e^{-2\mu + 2\lambda} (\lambda')^2)\,.
\end{align}

For the electromagnetic variables we would like to remove the integrals of $\bm Y^M_e$ which arise due to the solution of the Gau{\ss} constraint. For this we take the canonical transformation
\begin{align}
    A &:= - \int \bm Y^M_e \dd{r}\\
    \Pi_A &:=- \partial_r \bm X_M^{e}\,.
\end{align}

We now apply the transformation to the first order Hamiltonian constraint ${}^{(1)} Z^v_{lm}$. With the help of Mathematica we solve this constraint for $q_2$ and obtain
\begin{align}
    q^{(1)}_2 &=\frac{(\lambda')^2}{2 l (l+1) \Lambda}\Big[4 e^{4\lambda -4 \mu} (\Lambda + 2 \Delta)q_1 - e^{-3 \mu + 5 \lambda} (M_1 q_1)' + 2 \sqrt{l(l+1)} g^2 \xi A e^{4 \mu}\Big]\,.
\end{align}
In the above formula we used the gauge $p_2 =0$ which we will use from now on. 
This means that in this section we work in the gauge $\bm X^e=\bm x^e=p_2=0$. The true degrees of freedom are therefore $q_1,p_1$ and $\bm X^e_M, \bm Y_e^M$. This completes the solution of the first order non-symmetric constraints.

The canonical transformation and the solution of the first order constraints $Z^v$, $Z^h$ and $Z^e$ are inserted in the even parity contributions to $\pi_\mu^{(2)}$ in equation  \eqref{eq:SolIntegralpimu2}. 
The integral then only depends on the true degrees of freedom $q_1, p_1$ 
and $A, \Pi_A$. The form of the Hamiltonian is not particularly tractable. Thus, we perform two more 
canonical transformations to simplify it. First, we use
\begin{align}
    p_1 &= \sqrt{\frac{(l+2)(l-1)}{l(l+1)}}\frac{e^{-\mu + 2 \lambda}\lambda' M_1}{4 \Lambda}\qty(P + \frac{\xi e^{-\lambda}}{\sqrt{(l+2)(l-1)}}\Pi_A + A_3 A)\\
    q_1&=\sqrt{\frac{l(l+1)}{(l+2)(l-1)}}\frac{4 \Lambda}{e^{-\mu + 2 \lambda}\lambda' M_1}Q \\
    A &= \tilde A - \frac{\xi e^{-\lambda}}{\sqrt{(l+2)(l-1)}} Q\\
    \Pi_A &=\tilde \Pi_A + \Gamma Q
\end{align}
with the function $\Gamma$ defined as
\begin{align}
    \Gamma &= \frac{g^2 \xi}{2 e^{3 \mu + 6 \lambda} \lambda' \Lambda \sqrt{(l+2)(l-1)} \pi_\mu M_1}  \Big(\frac{e^{6 \lambda + 4 \mu}}{(\lambda')}\pdv{r}\left(16 \Lambda  e^{2
   \lambda - 4 \mu} (\lambda')^2 \left((\lambda'^2-\Delta  e^{2 \mu-2 \lambda}\right)\right)\\
   &+16 l(l+1) e^{4 (\lambda + \mu)} \left(e^{4 \lambda - 4 \mu} (\lambda')^4-\Delta^2\right)+16 \Lambda  l(l+1) e^{4 (\lambda + \mu)} \left(e^{2 \lambda - 2 \mu} (\lambda')^2-\Delta \right)\\
   &-(\lambda')^2
   \left(2 e^{2 (\lambda + \mu)} \left(-8 g^4 \xi^4 - 4 e^{2 \lambda}
   \left(8 g^2 \xi ^2+21 r_s^2\right)+53 g^2 \xi^2 r_s e^{\lambda}+84 r_s e^{3 \lambda}\right)\right)\\
   &-8 e^{6 \lambda} (\lambda')^4 \left(8 g^2
   \xi^2+4 e^{2 \lambda}-21 r_s e^{\lambda}\right) + 32 \Delta^2 e^{4 (\lambda + \mu
   )}\Big)\,.
\end{align}
This transformation scales $p_1$ and $q_1$ in the correct way. The shift in the transformation of $p_1$ leads to a removal of the coupling between the electromagnetic degrees of freedom $\tilde A,\tilde \Pi_A$ and $P$. The remaining coupling term is of the form $Q \tilde A$. 

The last transformation removes the terms proportional to $QP$ and $\tilde A \tilde \Pi_A$. This is achieved through a shift of the momentum variables:
\begin{align}
    Q &= Q^e, \quad \quad \quad ~~~~~P = P^e + A_{\mathrm{grav}} Q^e\\
    \tilde A &= \frac{1}{g^2} A^e, \quad \quad \quad \tilde \Pi_A = g^2 \Pi_A^e + A_{em} \frac{1}{g^2}A^e\,,
\end{align}
where
\begin{align}
    A_{\mathrm{em}} = -\frac{g^4 \xi^2 e^{-4 \lambda} \pi _{\mu }}{8 \Lambda}
\end{align}
The expression for the shift $A_{\mathrm{grav}}$ in the gravitational sector is very long. It is shown explicitly in appendix \ref{sec:EvenParityAnalysis}. 

The last step is to explicitly compute the even parity contributions to the solution for $\pi_\mu^{(2)}$. For this we use the solution of the first order constraints $Z^v$, $Z^h$ and $Z^e$ and all the canonical transformations mentioned in this subsection. The result of the computation will be an integral over the true degrees of freedom $Q^e, P^e$ and $A^e,\Pi_A^e$. The expression is simplified using integration by parts. The boundary term is recorded in appendix \ref{sec:EvenParityAnalysis}. We neglect it for now and will show later that this is justified. With the help of Mathematica we find
\begin{align}
     &\pi_\mu^{(2)}\Big|_{\mathrm{even}} = \frac{e^{\lambda+2\mu}}{\pi \pi_\mu^{(0)}} \int \dd{r}\Bigg[\frac{\pi_\mu^{(0)}}{4}   e^{-2 \mu - \lambda} (P^e \cdot (Q^e)' + \Pi_A^e \cdot  (A^e)')+ \frac{1}{2} e^{-2\mu + \lambda}\lambda' \qty(P^e \dot P^e + {Q^e}'\cdot Q^e{}' + V_\mathrm{grav}^{e} Q^e\cdot Q^e)\nonumber\\
     &+ \frac{1}{2}e^{-2\mu + \lambda}\lambda'\qty(g^2\Pi_A^e \cdot \Pi_A^e + \frac{1}{g^2} A^e{}' \cdot A^e{}' + \frac{1}{g^2} V_{\mathrm{em}}^{e} A^e\cdot A^e) + e^{-2\mu + \lambda}\lambda' \frac{1}{g} V_\mathrm{coup}^{e} Q^e\cdot A^e\Bigg]\,.
\end{align}
In the result we introduced the gravitational potential $V^e_{\mathrm{grav}}$, the electromagnetic potential $V^e_{\mathrm{em}}$ and the potential for the coupling term $V^e_{\mathrm{coup}}$. They are defined analogously to the odd parity case as 
\begin{align}
    V_{\mathrm{coup}}^{e} &:=e^{2 \mu - 3 \lambda} g \xi W^e\\
    V_{\mathrm{grav}}^{e} &:= e^{2 \mu - 2\lambda}\qty(U^e - \frac{3}{2} e^{-\lambda} r_s W^e)\\
    V_{\mathrm{em}}^{e} &:= e^{2\mu - 2\lambda}\qty(U^e + \frac{3}{2} e^{-\lambda} r_s W^e)\,.
\end{align}
The difference between the potentials in the odd and even parity sectors is the functions $U^e$ and $W^e$. They are more complicated in the even parity sector and are given by
\begin{align}
    W^{e} &:= \frac{\Delta}{\Lambda^2} \qty(2 n + \frac{3}{2} e^{-\lambda} r_s) + \frac{1}{\Lambda}\qty(n + \frac{1}{2}e^{-\lambda} r_s)\\
    U^{e} &:= \qty(2 n  + \frac{3}{2} e^{-\lambda} r_s) W + \qty(\Lambda - n - \frac{1}{2}e^{-\lambda} r_s) - \frac{2 n \Delta}{\Lambda}
\end{align}

We now consider the boundary term in appendix \ref{sec:EvenParityAnalysis}. For the analysis we use the asymptotic expansion found in equations \eqref{eq:Decay} and \eqref{eq:FAllOffEM}. They imply the following asymptotic behaviour of the variables defined in this section:
\begin{align}
   q_1 &\sim q_1^0 r^{-1} & p_1 &\sim p_1^0 & q_2 &\sim q_2^0 r & A &\sim A_0 r & A^e&\sim A^e_0 r & Q^e &\sim Q^e_0 \,.
\end{align}
The quantities with sub-/superscript $0$ are radial constants but are allowed to vary with respect to $l,m$. Furthermore, we use the asymptotics of the background degrees of freedom $\mu,\lambda$ from equation \eqref{eq:FallOffBackground}. Using these definitions the boundary term of $\pi_\mu^{(2)}$ behaves as
\begin{align}
    \begin{split}
    &\frac{1}{\pi \pi_\mu^\infty \sqrt{r}}\Big(- \frac{1}{2}(p_1^0)^2 + \frac{3}{2} (q_1^0)^2 + \frac{(l^2 + l + 2)}{(l+2)(l+1)l(l-1)} (q_1^0)^2 + 2 q_1^0 q_2^0 - \frac{3(l^2+l+2)}{2} q_1^0 q_2^0 + \frac{1}{2} (q_2^0)^2\\
    &- (l^2 + l + 2) (q_2^0)^2+ \frac{1}{8}(3 l^4+ 6 l^3+13 l^2+10l + 16) (q_2^0)^2+ \frac{g^2(l^2+l+2) \xi}{\sqrt{l(l+1)}(l+2)(l-1)} A_0 q_1^0\\
    &- \frac{1}{2} g^2 \xi \sqrt{l(l+1)} q_2^0 A_0 - \frac{1}{2} (Q^e_0)^2 + \frac{g^4 \xi^2}{2(l+2)(l-1)} (A^e_0)^2 + 2 \frac{g^2 \xi}{2 \sqrt{(l+2)(l-1)}}A^e_0 Q^e_0\Big) + O(r^{-1})
    \end{split}
\end{align}
We observe that the leading contributions vanish like $r^{-1/2}$ in the limit $r\to \infty$. This shows that the boundary term is not relevant as $r$ goes to infinity and we are allowed to drop it in the calculation.

We worked in the gauge $\bm X^e=\bm x^e = p_2=0$. This is different from the gauge used in the other papers of this series where we set only $x=0$. The reason for our choice of gauge is that we do not have to solve any differential equation in the analysis. Furthermore, we do not need to handle integrals of the degrees of freedom in the computation.

In \cite{II} we showed that the formulation in different gauges for the perturbations is equivalent provided that the physical Hamiltonian in the end is weakly gauge invariant. In the previous papers we could show that the problematic gauge-variant contributions are equal to a boundary term which vanishes in the limit $r$ to infinity. We could then relate the true degrees of freedom in the two approaches and are guaranteed to obtain the same result. We believe that also here the physical Hamiltonian is weakly gauge invariant and that the obstruction terms vanish at infinity.

It would be desirable to follow the same steps as in \cite{II,III} and to proof that the physical Hamiltonian is weakly gauge invariant.
However already in the previous papers the boundary terms were lengthy and it is expected that they will get even longer in a general gauge for the background. For this reason we will not study the problem here and leave it open for future research.

\subsection{Decoupling of the Equations}

In the previous paper \cite{III} we showed that it is possible to decouple the expressions for the gravitational and electromagnetic degrees of freedom. In this paper we reduced the problem to similar expressions for $\pi_\mu^{(2)}$. In the following we use the same arguments to remove the coupling terms. First, we consider a canonical transformation parametrized by an angle $\theta$ as follows 
\begin{align}
    \mqty(Q^{o/e} \\ A^{e/o}) &= \mqty( \cos \theta & \frac{1}{g} \sin \theta\\ - g \sin \theta & \cos \theta) \mqty( Q_1^{e/o} \\ Q_2^{e/o})\,, & \mqty(P^{e/o} \\ \Pi_A^{e/o}) &= \mqty(\cos \theta & g \sin \theta \\ - \frac{1}{g} \sin \theta & \cos \theta) \mqty(P_1^{e/o} \\ P_2^{e/o})\,.
\end{align}
In the last paper we showed that there is a special choice for the angle $\theta$ such that the coupling terms in the transformed variables vanish. The same choice for $\theta$ works here as well because the potentials $V^{e/o}_\mathrm{grav}$, $V^{e/o}_\mathrm{em}$ and $V^{e/o}_\mathrm{coup}$ have the same form. In \cite{III} we found that $\theta$ is defined implicitly as
\begin{align}
    \cos(2\theta) &= \frac{3 r_s}{\sqrt{9 r_s^2 + 4 (l+2)(l-1) g^2 \xi^2}}\,,\\
    \sin(2\theta) &= \frac{2 \sqrt{(l+2)(l-1)} g \xi}{\sqrt{9 r_s^2 + 4 (l+2)(l-1) g^2 \xi^2}}\,.
\end{align}
Applying the canonical transformation to the variables $Q_1,P_1$ and $Q_2,P_2$ the contributions to $\pi_\mu^{(2)}$ for both the odd and the even parity sector are
\begin{align}
    \pi_\mu^{(2)} = &\sum_{I \in \{e,o\}}\frac{e^{\lambda + 2 \mu}}{\pi \pi_\mu^{(0)}} \int \dd{r}\Bigg[ \frac{1}{4}e^{-\lambda - 2 \mu} \pi_{\mu}^{(0)}(P^I_1\cdot  Q^I_1{}' + P^I_2 \cdot Q^I_2{}')\\
    &+ \frac{1}{2} e^{-2\mu + \lambda} \lambda{}'\Big( P^I_1 \cdot P^I_1 + g^2 P^I_2 \cdot P^I_2  + Q^I_1{}'\cdot Q^I_1{}' + \frac{1}{g^2}Q^I_2{}' \cdot Q^I_2{}' + V^I_1 Q^I_1\cdot Q^I_1 + \frac{1}{g^2}V^I_2 Q^I_2 \cdot Q^I_2\Big)\Bigg]\,,\nonumber
\end{align}
In the expression we have the potentials
\begin{align}
    V_1^I &= e^{2\mu -2 \lambda}\qty(U^I - \frac{1}{2} e^{-\lambda} \sqrt{9 r_s^2 + 4 (l+2)(l-1) g^2 \xi^2} W^I)\\
    V_2^I &= e^{2\mu -2 \lambda}\qty(U^I + \frac{1}{2} e^{-\lambda} \sqrt{9 r_s^2 + 4 (l+2)(l-1) g^2 \xi^2} W^I)\,,
\end{align}
They are constructed from $U^{e/o}$ and $W^{e/o}$ corresponding to the value of $I$.


\subsection{The physical Hamiltonian}

The final step of the program is the derivation of the physical Hamiltonian. 
%
%
In \cite{I} it is derived in the Gullstrand-Painlevé gauge. The non-perturbative calculation gives
\begin{align}
    H = \lim_{r \to \infty} \frac{\pi}{2\kappa r}\pi_\mu^2\,,
    \label{eq:PhysHam}
\end{align}
where $\kappa = 16 \pi$ is the gravitational coupling constant. In the following we argue that the same expression also applies to the generalised gauges we discuss here. 
The derivation of the physical Hamiltonian needs two ingredients, the boundary term and the solution of the stability condition of the gauge fixing. 
In this paper we chose the fall-off conditions on the canonical variables to be compatible with the Gullstrand-Painlevé gauge.
Therefore, the boundary term we have to add to the theory is the same as in \cite{I}.
In the notation used in this paper it is given by $N^3 \pi_\mu$. 
It remains to explicitly calculate the asymptotic behaviour of the shift $N^3$ given some arbitrary gauge fixing $\mu = \mu^*(r)$ and $\lambda = \lambda^*(r)$ compatible with \eqref{eq:FallOffBackground}. 

For the calculation we recall the gravitational part of the Hamiltonian constraint $V_0$ and diffeomorphism constraint $V_\mu$. This is sufficient for the computation because the matter contributions to the constraints do not depend on the gravitational momenta $\pi_\mu$ and $\pi_\lambda$. The constraints of the full theory are given by
\begin{align}
    V_0^{\mathrm{grav}} &= \frac{1}{\sqrt{m}}\qty(m_{\mu \rho} m_{\nu \sigma} - \frac{1}{2} m_{\mu \nu} m_{\rho \sigma})W^{\mu \nu} W^{\rho \sigma} - \sqrt{m} {}^{(3)}R\\
    V_\mu^{\mathrm{grav}} &= - 2 m_{\mu \nu} \nabla_{\rho} W^{\nu \rho}
\end{align}

We introduce the smeared version of these constraints as $V_0[S] = \int \dd \Sigma S V_0$ and $\vec V[\vec S] = \int \dd \Sigma S^\mu V_\mu$. The stability condition is then evaluated by requiring that the Poisson bracket of the Hamiltonian with the variables $\mu$ and $\lambda$ vanishes:
\begin{align}
    \qty{\mu, V_0[S] + \vec V[\vec S]} &= \int_{S^2} \dd \Omega \sqrt{\Omega}\Big[\frac{S}{\sqrt{m}}\qty(m_{3\mu} m_{3 \nu} - \frac{1}{2} m_{33} m_{\mu \nu})e^{-2 \mu}  W^{\mu \nu} + e^{-2\mu} m_{\mu 3} \nabla_3(S^\mu)\Big]=0\\
    \begin{split}
    \qty{\lambda, V_0[S] + \vec V[\vec S]} &= \int_{S^2} \dd{\Omega} \sqrt{\Omega}\Big[\frac{S}{2\sqrt{m}}\qty(m_{A\mu} m_{B \nu} - \frac{1}{2} m_{AB} m_{\mu \nu})e^{-2 \lambda} \Omega^{AB} W^{\mu \nu}\\
    &~~~~~~~~~~~~~~~~~~+ \frac{1}{2} e^{-2\lambda}\Omega^{AB}m_{\mu A} \nabla_B(S^\mu)\Big] =0
    \end{split}
\end{align}
In the calculation we used the explicit form for the metric $m_{\mu \nu}$ and conjugate momentum $W^{\mu \nu}$ in terms of the background variables. 

The full set of these equations is very difficult to solve. However, we are only interested in the solution of the shift in the limit as $r$ approaches infinity. Thus, we solve the equations asymptotically for large $r$. In the following we will drop subleading terms from the equations. The behaviour of background and perturbations at infinity implies the following asymptotic behaviour of the terms proportional to $S$:
\begin{align}
    \qty{\mu, V_0[S] + \vec V[\vec S]} &\sim \int_{S^2} \dd \Omega \sqrt{\Omega} \Big[\frac{S}{4 \sqrt{m}}\qty(\pi_\mu^\infty - \pi_\lambda^\infty)\sqrt{r} + m_{3\mu} \partial_3(S^\mu) + \Gamma_{3 3 \nu} S^\nu\Big]=0\\
    \qty{\lambda, V_0[S] + \vec V[\vec S]} &\sim \int_{S^2} \dd{\Omega} \sqrt{\Omega} \Big[- \frac{S}{4\sqrt{ m}} \pi_\mu^\infty \sqrt{r} + \frac{1}{2} r^{-2}\Omega^{AB}m_{\mu A} \partial_B(S^\mu)  + \frac{1}{2}r^{-2}\Omega^{AB} \Gamma_{A B\nu}(S^\nu)\Big]=0
\end{align}
In the above equations we expanded the terms involving $S^\mu$ into Christoffel symbols. The asymptotic behaviour of the Christoffel symbols is
\begin{align}
\begin{split}
    \Gamma_{333} &\sim - \frac{1}{2r^2} \delta m_{33}^+\\
    \Gamma_{33A} &\sim \frac{1}{2r}\partial_A \delta m_{33}^+\\
    \Gamma_{3AB} &\sim \frac{1}{2}\qty(\partial_B \delta m_{3A}^+ + \partial_A \delta m_{3B}^+ - 2 r \Omega_{AB}) \sim - r \Omega_{AB}\\
    \Gamma_{A33} &\sim - \frac{1}{2r}\partial_A \delta m_{33}^+\\
    \Gamma_{AB3} &\sim \frac{1}{2}\qty(\partial_B \delta m_{3 A}^+ - \partial_A \delta m_{3B}^+ + 2 r \Omega_{AB}) \sim r \Omega_{AB}\\
    \Gamma_{ABC} &\sim \frac{r^2}{2}\Gamma_{ABC}^\Omega
\end{split}
\end{align}
where $\Gamma_{ABC}^\Omega$ is the Christoffel symbol on the sphere. For the square root of the determinant of the metric we write down $m$ explicitly and compute the determinant of the $3\times3$ matrix. The computation gives the leading order behaviour $\sqrt{m} \sim r^2 \sqrt{\Omega}$. The expressions for the Christoffel symbols and the expansion for $\sqrt{m}$ drastically simplify the equations that we have to solve. If we further assume that $S^A = O(r^{-2})$ we obtain
\begin{align}
    \qty{\mu, V_0[S] + \vec V[\vec S]} &\sim 4 \pi \Big[\frac{N}{4 r^{3/2}}\qty(\pi_\mu^\infty - \pi_\lambda^\infty) + \partial_3(N^3)\Big]=0\\
    \qty{\lambda, V_0[N] + \vec V[\vec N]} &\sim  4\pi \Big[- \frac{N}{4 r^{3/2}} \pi_\mu^\infty + r^{-1} N^3\Big]=0
\end{align}
In the equations the integral over the sphere reduces the quantities $S$, $S^3$ to their symmetric parts $N$ and $N^3$ respectively.

We now solve the differential equations for $N$ and $N^3$. The second equation gives a relation between $N$ and $N^3$:
\begin{equation}    
    N = \frac{4 \sqrt{r}}{\pi_\mu^\infty} N^3
\end{equation}
Replacing $N$ in the first equation gives a differential equation for $N^3$:
\begin{align}
    \partial_3 N^3 + \frac{1}{r}\qty(1 - \frac{\pi_\lambda^\infty}{\pi_\mu^\infty}) N^3=0
\end{align}

The constants $\pi_\mu^\infty$ and $\pi_\lambda^\infty$ are not independent. The relation between the two is fixed by looking at the radial diffeomorphism constraint. In its general form it is given by
\begin{align}
    V_\mu = - 2 m_{\mu \nu} \nabla_\rho W^{\nu \rho} - 2 \Gamma_{\mu \nu \rho} W^{\nu \rho}.
\end{align}
We insert the background and the perturbations in this constraint and consider its asymptotic behaviour near infinity. The dominant contribution behaves as $r^{-1/2}$ and gives $\pi_\mu^\infty = 2 \pi_\lambda^\infty$. Therefore the equation for $N^3$ becomes
\begin{align}
    \partial_3 N^3 + \frac{1}{2r} N^3 = 0\,.
\end{align}
The solution of this equation is
\begin{align}
    N^3 = \frac{C\pi_\mu^\infty}{4\sqrt{r}}\,,
\end{align}
with some integration constant $C$. For this form of the shift the lapse function becomes
\begin{equation}
    N = C
\end{equation}
We observe that for $C=1$ the lapse asymptotes to unity and we will use this value for $C$ from now on.

It is convenient to rewrite the shift $N^3$ in the limit as $r$ approaches infinity. The above formula for $N^3$ is equivalent to
\begin{equation}
    N^3 = \frac{\pi_\mu}{4 r}\,,
\end{equation}
This is precisely the result obtained in the Gullstrand - Painlevé gauge. Therefore, the formula for the physical Hamiltonian is still applicable for the generalised gauges discussed in this paper.

For the interpretation of the result later it is advantageous to explicitly compute the full solution of the zeroth order stability condition. 
The condition for $N_{(0)}$ and $N^3_{(0)}$ reads
\begin{align}    
    \dot \mu &= 4 \pi\qty[\frac{1}{4}e^{-\mu - 2\lambda}(\pi^{(0)}_\mu - \pi^{(0)}_\lambda) N_{(0)} + (N^3_{(0)})' + N^3_{(0)} \mu']=0\,,\\
    \dot \lambda &= 4 \pi\qty[- \frac{1}{4} e^{-\mu - 2 \lambda} \pi_\mu^{(0)} N_{(0)} + N^3_{(0)} \lambda']=0\,,
\end{align}
where $\pi_\mu^{(0)}$ and $\pi_\lambda^{(0)}$ need to be replaced by the corresponding solution of the constraint equations. The solution to these equations is given by
\begin{align}
    N_{(0)} &= \lambda' e^{-\mu +\lambda}\,,\\
    N^3_{(0)} &= \frac{\pi_\mu}{4}e^{-\lambda - 2 \mu}\,.
\end{align}
In the integration of the differential equation we obtained an integration constant. 
This constant was chosen such that the lapse function asymptotes to one.
We observe that the shift behaves like $N^3 \sim \flatfrac{\pi^\infty_\mu}{(4\sqrt{r})}$ at infinity. This is fully compatible with the results above and the  behaviour obtained in \cite{I}. 

The physical Hamiltonian to second order as defined in equation \eqref{eq:PhysHam} is given by 
\begin{align}
\begin{split}
    H=\lim_{r\to \infty} \frac{\pi}{2 \kappa r}\pi_\mu^2 &= \lim_{r\to \infty} \frac{\pi}{2\kappa r}\qty((\pi_\mu^{(0)})^2 + 2 \pi_\mu^{(0)} \pi_\mu^{(2)} + O(3))\\
    &= M + \frac{1}{\kappa} \int_{\mathbb{R}^+} \dd{r}\Big[ N^3_{(0)} {}^{(2)}C_h + N_{(0)} {}^{(2)}C_v\Big] + O(3)\,.
\end{split}
\end{align}
We interpreted the prefactors of ${}^{(2)}C_h$ and ${}^{(2)}C_v$ as the lapse and shift obtained from solving the zeroth order stability conditions.

To summarise we have the physical Hamiltonian
\begin{equation}
    H = M + \frac{1}{\kappa} \sum_{l\geq2,m,I} \int_{\mathbb{R}^+} \dd{r}\Big[ N^3_{(0)} P^I_{lm} \partial_r Q^I_{lm} + \frac{N_{(0)}}{2} e^{-\mu} \qty((P^I_{lm})^2 + (\partial_r Q^I_{lm})^2 + V_I (Q^I_{lm})^2)\Big] + O(3)\,,
\end{equation}
where we restored the labels $l$ and $m$. $I$ stands for the labels even and odd as well as 1 and 2 from the previous chapter. The potentials $V_I$ are the decoupled potentials.

\section{Relation to Lagrangian Formalism:}
\label{sec:Comparison}

We finish the paper by showing that the equations of motion of the physical Hamiltonian agree with those found in the literature (see e.g. \cite{Chandrasekhar1983}). 
The second order contributions to the physical Hamiltonian were all brought to a similar form, a Hamiltonian of a scalar field with a mass term depending on the background true degrees of freedom, the electric charge $\xi$ and the mass $M$.  The general form of the second order contributions is 
\begin{align}
    \int_{\mathbb{R}^+} \dd{r} N^3 P Q' + \frac{N}{2}e^{-\mu}\qty(P^2 + (Q')^2 + V Q^2)\,.
\end{align}
Here $Q,P$ are some generic position and momentum variables and $V$ is the appropriate potential. For simplicity we do not display the subscript ${(0)}$ on lapse $N$ and shift $N^3$. In the following we study the Hamiltonian equations of motion and compare them to the equations of motion found in the literature. We have
\begin{align}
    \dot Q &= N^3 Q' + N e^{-\mu} P\,,\\
    \dot P &= \partial_r (N^3 P) + \partial_r (N e^{-\mu}Q') - N e^{-\mu} V Q\,.
\end{align}
The system of two first order equations can be combined into a second order differential equation. For this we solve the first equation for $P$ and insert the result into the second equation. This gives a second order differential equation for $Q$:
\begin{align}
    - \partial_t \qty(\frac{e^\mu}{N}\qty(\dot Q - N^3 Q')) + \partial_r \qty(\frac{e^\mu N^3}{N}\qty(\dot Q - N^3 Q')) + \partial_r (N e^{-\mu}Q') - N e^{-\mu} V Q =0\,.
    \label{eq:DiffEqnQ1}
\end{align}
It is possible to formulate the equation in a covariant form by introducing the wave operator $\square$ on the $(t,r)$ hypersurfaces. We denote the metric on these hypersurfaces by $g$. The metric and its inverse $g^{-1}$ are given by
\begin{align}
    g = \mqty(-N^2 + e^{2\mu}(N^3)^2 & e^{2\mu} N^3\\ e^{2\mu} N^3 & e^{2\mu}), ~~~~~~~~ g^{-1} = \mqty(-\frac{1}{N^2} & \frac{N^3}{N^2}\\ \frac{N^3}{N^2} & e^{-2\mu} - \frac{(N^3)^2}{N^2})\,,
\end{align}
with $\sqrt{-\det(g)}= N e^{\mu}$. Therefore dividing equation \eqref{eq:DiffEqnQ1} by $Ne^{\mu}$ and rearranging the terms we get
\begin{align}
    \frac{1}{\sqrt{-g}}\partial_t\qty(\sqrt{-g}\qty(g^{tt} \dot Q + g^{t3} Q')) + \frac{1}{\sqrt{-g}}\partial_r \qty(\sqrt{-g}\qty(g^{3t} \dot Q + g^{33} Q')) = e^{-2\mu} V Q\,.
\end{align}

The differential operator on the left hand side is just the Laplace operator $\square$ associated to $g$. Hence, the differential equation is equivalent to a wave equation for $Q$ with potential $V$:
\begin{equation}
    \square Q = e^{-2\mu} V Q\,.
\end{equation}
The difference between the different sectors and the electric and gravitational degrees of freedom is the potential which shows up on the right-hand side. 
We compare the potentials with the ones obtained in the literature.
The results in \cite{Chandrasekhar1983} are derived in the Schwarzschild gauge and we reduce our results to this gauge.
Since the equation is covariant this is done by setting $e^{-2\mu}= \Delta$ and $e^\lambda=r$. We find that our potentials match precisely the potentials derived by Chandrasekhar.

\section{Conclusion}
\label{sec:Conclusion}

We studied second order corrections to spherical symmetric general relativity coupled to electromagnetism. 
In contrast to the previous publications of the series \cite{II,III} we did not restrict to a specific gauge for the background degrees of freedom.
We show that in the general setup the same methods produce a physical Hamiltonian which is consistent with the previous results. The equations of motion of the final Hamiltonian match the results discussed in the literature.

In the future, extending our approach to gravity coupled to other matter fields is of particular interest. 
Specifically, incorporating fermions, such as neutrinos, presents an important generalization. 
Fermions are a major part of the Standard Model of particle physics. Therefore, understanding their dynamics 
in the vicinity of black holes is crucial for a complete theory of black holes.

\section*{Acknowledgements}
The author thanks the Hanns-Seidel-Stiftung for financial and intellectual support and Thomas Thiemann for insightful discussions throughout the project. 

\appendix

\section{Expansion of the Hamiltonian and Diffeomorphism Constraints to Second Order}
\label{sec:Expansion2ndOrder}

In this section we show some details of the derivation of the perturbative expansion of the Hamiltonian and diffeomorphism constraints. It is based on a similar appendix in \cite{II} where we worked in the Gullstrand-Painlevé gauge. Here we are not fixing the background degrees of freedom to any specific choice of gauge. For this general setup the background Christoffel symbols take the form
\begin{align}
    \Gamma^3_{33} &= \mu' \\
    \Gamma^3_{AB} &= - \lambda' e^{2\qty(\lambda - \mu)} \Omega_{AB}\\
    \Gamma^A_{3B} &= \lambda'\delta^A_B,
\end{align}
$\Gamma^A_{BC}$ is the Christoffel symbol on the 2-sphere. All other components of the Christoffel symbol vanish. The components of the curvature tensors for the background are
\begin{align}
\begin{split}
    R_{33} &= - 2 \qty(\lambda'' + \lambda'(\lambda' - \mu')) \\
    R_{AB} &= \Omega_{AB} \qty(1 - (\lambda')^2 e^{-2 \mu + 2 \lambda} - \qty(\lambda'' + \lambda'(\lambda' - \mu')) e^{2(\lambda - \mu)}) \\
    R &= 2 e^{-2 \lambda} - 2 (\lambda')^2 e^{-2 \mu} - 4 \qty(\lambda'' + \lambda' (\lambda' - \mu')) e^{-2 \mu}\\
    G_{33} =& (\lambda')^2 - e^{2 (\mu -  \lambda)} \\
    G_{AB} =& \Omega_{AB} \qty(\lambda'' + \lambda'(\lambda' - \mu')) e^{2(\lambda - \mu)}
\end{split}
\end{align}

For the computation we split the induced metric $m_{\mu \nu}$ into background and perturbations. We define $m_{\mu \nu} = \overline m_{\mu \nu} + \delta m_{\mu \nu}$ and $W^{\mu \nu} = \overline W^{\mu \nu} + \delta W^{\mu \nu}$. The bared quantities are the background given in terms of $\mu, \pi_\mu$ and $\lambda, \pi_\lambda$ as shown in equation \eqref{eq:DefSymVars}. The variables with $\delta$ are the perturbations. The indices are raised and lowered with the background metric $\overline m_{\mu \nu}$. The formulas for the perturbations of the inverse metric, the Riemann curvature and the Ricci scalar are found in \cite{II}.

First we consider the gravitational contributions to the constraints. We start with perturbing the diffeomorphism constraints. To first order we find
\begin{equation}
    {}^{(1)}V^\mathrm{grav}_\mu = - 2 \nabla_\rho \qty(\delta m_{\mu\nu} \overline W^{\nu \rho} + \overline m_{\mu \nu} \delta W^{\nu \rho}) + \overline W^{\nu \rho} \nabla_\mu \delta m_{\nu \rho}\,.
\end{equation}
The indices $\mu,\nu,\dots$ are split into the radial components $3$ and angular components $A,B,\dots$. For the first order diffeomorphism constraint we find
\begin{align}
\begin{split}
    {}^{(1)}V^\mathrm{grav}_3 =& -2 e^{\mu} \partial_r \qty(e^\mu \delta W^{33})- 2 e^{2\mu} D_A \delta W^{3A} + 2  e^{2 \lambda} \lambda' \Omega_{AB} \delta W^{AB} - \sqrt{\Omega} \partial_r(\pi_\mu e^{-2 \mu}) \delta m_{33} \\
    &- \sqrt{\Omega}\frac{\pi_\mu}{2} e^{-2 \mu} \partial_r \delta m_{33} - \sqrt{\Omega} \frac{\pi_\lambda}{2} e^{-2 \lambda} D^A \delta m_{3A} + \sqrt{\Omega}\frac{\pi_\lambda}{4} e^{-2\lambda} \Omega^{AB} \partial_r \delta m_{AB} \,,
\end{split}\\
\begin{split}
    {}^{(1)}V^\mathrm{grav}_A =& -2 \Omega_{AB} \partial_r \qty(e^{2\lambda} \delta W^{B3}) - 2 e^{2\lambda}\Omega_{AB} D_C \delta W^{BC}+ \frac{\sqrt{\Omega} \pi_\mu}{2} e^{-2\mu} D_A \delta m_{33}\\
    &- \sqrt{\Omega} \partial_r \qty(\pi_\mu e^{-2\mu}\delta m_{3A} )+ \frac{\sqrt{\Omega} \pi_\lambda}{4}e^{-2\lambda} \qty(D_A \Omega^{CD} \delta m_{CD}-2 D^B \delta m_{AB})\,.
\end{split}
\end{align}
To second order the diffeomorphism constraint takes the form
\begin{equation}
    {}^{(2)}V^\mathrm{grav}_\mu = - 2 \nabla_\rho \qty(\delta m_{\mu \nu} \delta W^{\nu \rho}) + \delta W^{\nu \rho} \nabla_\mu \delta m_{\nu \rho} \,.
\end{equation}
For the symmetric constraints to second order we only need the radial component $\mu=3$. It is given by
\begin{align}
\begin{split}
    {}^{(2)}V^\mathrm{grav}_3 =& \delta W^{33} \partial_r \delta m_{33} - 2 \partial_r \qty(\delta W^{33} \delta m_{33}) - 2 \delta m_{3A} \partial_r \delta W^{3A}\\
    &- 2 D_A \qty(\delta m_{33} \delta W^{3A}) + \delta W^{AB} \partial_r \delta m_{AB} - 2 D_A \qty(\delta m_{3B} \delta W^{AB})\,.
\end{split}
\end{align}
The first order perturbation of the Hamiltonian constraint is given by
\begin{align}
    \begin{split}
    {}^{(1)}V^\mathrm{grav}_0 =&\frac{1}{\sqrt{\overline m}}\qty(\overline W^{\rho \sigma} \overline m_{\mu \rho} \overline m_{\nu \sigma} - \frac{1}{2} \overline m_{\mu \nu} \overline W^{\rho \sigma}\overline m_{\rho \sigma})\qty(2 \delta W^{\mu \nu}+ 2 \overline m^{\mu \alpha}  \delta m_{\alpha \beta} \overline W^{\beta \nu}- \frac{1}{2} \overline m^{\alpha \beta} \delta m_{\alpha \beta} \overline W^{\mu \nu})\\
    & + \sqrt{\overline m}\qty(\overline G_{\mu \nu} \delta m_{\rho \sigma} \overline m^{\mu\rho} \overline m^{\nu \sigma}- \nabla_\mu \nabla_\nu \delta m_{\rho \sigma} \overline m^{\mu \rho} \overline m^{\nu \sigma} + \square \delta m_{\mu \nu} \overline m^{\mu \nu})\,.
    \end{split}
\end{align}
The splitting of the tensor indices into angular and radial components gives
\begin{align}
\begin{split}
    {}^{(1)}V^\mathrm{grav}_0 =& \frac{1}{2}\qty(\pi_\mu - \pi_\lambda) e^{\mu - 2 \lambda} \delta W^{33} - \frac{1}{2} \pi_\mu e^{- \mu}  \delta W^{AB} \Omega_{AB} - \frac{1}{16} \sqrt{\Omega} \pi_\mu^2 e^{- \mu - 4 \lambda} \Omega^{AB} \delta m_{AB}\\
    & + \frac{3}{16} \sqrt{\Omega} \pi_\mu^2 e^{- 3 \mu - 2 \lambda} \delta m_{33} - \frac{1}{8} \sqrt{\Omega} \pi_\mu \pi_\lambda e^{- 3 \mu - 2 \lambda} \delta m_{33}\\
    &+e^{\mu+ 2\lambda}\Big[e^{-2(\mu + \lambda)} ( D_A D^A \delta m_{33} -  \delta m_{33})+ e^{-2(\mu + \lambda)} \qty( \partial_r^2 - \mu' \partial_r - \lambda' \partial_r + (\lambda')^2) \Omega^{AB} \delta m_{AB}\\
    &+ e^{-4 \mu}\qty(- 2 \lambda' \partial_r  + 6 \lambda' \mu'  - 3 (\lambda')^2  - 2 \lambda'' ) \delta m_{33} -2 e^{-2 (\mu + \lambda)}(\partial_r - \mu' + \lambda') D^A \delta m_{3A}\\
    &+ e^{-4\lambda}\qty(D_A D^A \delta m_{CD} \Omega^{CD} - D^A D^B \delta m_{AB})\Big]\,.
\end{split}
\end{align}
For the second order contributions to the Hamiltonian constraint we have
\begin{align}
    \label{eq:Ham2ndOrder}
    {}^{(2)}V_0^\mathrm{grav} =&\frac{1}{\sqrt{\overline{m}}}\Big[\delta W^{\mu \nu} \delta W^{\rho \sigma} \qty(\overline m_{\mu \rho} \overline m_{\nu \sigma} - \frac{1}{2} \overline m_{\mu \nu}\overline m_{\rho \sigma}) + \delta W^{\mu \nu} \overline W^{\rho \sigma} \qty(4 \delta m_{\mu \rho} \overline m_{\nu \sigma} - \qty(\delta m_{\mu \nu} \overline m_{\rho \sigma} + \delta m_{\rho \sigma} \overline m_{\mu \nu}))\nonumber\\
    &+\overline W^{\mu \nu} \overline W^{\rho \sigma}\qty(\delta m_{\mu \rho} \delta m_{\nu \sigma} - \frac{1}{2} \delta m_{\mu \nu} \delta m_{\rho \sigma}) + \frac{1}{8}\qty(2 \delta m^{\mu \nu} \delta m_{\mu \nu} + (\delta m^{\mu}{}_{\mu})^2 )\qty(\overline W^{\rho \sigma}\overline W_{\rho \sigma} - \frac{1}{2} (\overline W^\rho{}_\rho)^2)\nonumber\\
    &+ \frac{1}{2} \delta m^\rho_\rho\qty( \delta W^\mu{}_\mu W^\nu{}_\nu + W^{\sigma}{}_\sigma W^{\mu \nu}  \delta m_{\mu \nu} - 2 W^{\mu \nu}W^{\sigma \tau} \delta m_{\mu \sigma} m_{\nu \tau} -2 \delta W^{\mu \nu} W_{\mu \nu} )\Big]\\
    &- \sqrt{\overline m}\Big[\delta m^{\mu}{}_\rho \delta m^{\rho \nu} \overline R_{\mu \nu}  - \frac{1}{2} \delta m^\rho{}_\rho \delta m^{\mu\nu}\overline R_{\mu \nu} + \frac{1}{8} \qty((\delta m^\rho{}_\rho)^2 - 2 \delta m^{\mu \nu}\delta m_{\mu \nu}) \overline R \nonumber\\
    &~~~~~~~~~+ \nabla_\mu \delta m_{\nu \rho} \nabla_\sigma \delta m_{\alpha \beta} \overline m_{(1)}^{\mu \nu \rho \sigma \alpha \beta} + \delta m_{\mu \nu} \nabla_\rho \nabla_\sigma \delta m_{\alpha \beta} \overline m_{(2)}^{\mu \nu \rho \sigma \alpha \beta}\Big]\nonumber\,,
\end{align}
where we introduced the following tensors:
\begin{align}
    \overline m_{(1)}^{\mu \nu \rho \sigma \alpha \beta}:=&-\overline m^{\mu \nu} \overline m^{\rho \beta} \overline m^{\sigma \alpha} + \overline m^{\mu \nu} \overline m^{\rho \sigma} \overline m^{\alpha \beta} + \frac{3}{4} \overline m^{\mu \sigma} \overline m^{\nu \alpha} \overline m^{\rho \beta} - \frac{1}{4} \overline m^{\mu \sigma} \overline m^{\nu \rho} \overline m^{\alpha \beta} - \frac{1}{2} \overline m^{\mu \beta} \overline m^{\nu \alpha} \overline m^{\rho \sigma}\\
    \begin{split}
    \overline m_{(2)}^{\mu \nu \rho \sigma \alpha \beta} :=& \frac{1}{2} \overline m^{\mu \nu} \overline m^{\rho \alpha} \overline m^{\sigma \beta} - \frac{1}{2} \overline m^{\mu \nu} \overline m^{\rho \sigma} \overline m^{\alpha \beta} +  \overline m^{\mu \alpha}\overline m^{\nu \beta} \overline m^{\rho \sigma} + \overline m^{\mu \rho} \overline m^{\nu \sigma} \overline m^{\alpha \beta} - \overline m^{\mu \sigma} \overline m^{\nu \alpha} \overline m^{\rho \beta}\\
    &- \overline m^{\mu \rho} \overline m^{\nu \alpha} \overline m^{\sigma \beta}\,.
    \end{split}
\end{align}

For the computation it is convenient to expand the second covariant derivatives into the radial and angular components. For this we introduce the differential operator
\begin{equation}
    D^{(a,b)} = \partial_r - a \mu' - b \lambda',
\end{equation}

The second derivatives of $\delta m_{\mu \nu}$ in radial and angular components are
\begin{align}
    \nabla_r \nabla_r \delta m_{33} &= D^{(3,0)} D^{(2,0)}\delta m_{33}\nonumber\\
    \nabla_r \nabla_r \delta m_{3A} &= D^{(2,1)} D^{(1,1)} \delta m_{3A}\nonumber\\
    \nabla_r \nabla_r \delta m_{AB} &= D^{(1,2)}D^{(0,2)} \delta m_{AB}\nonumber\\
    \nabla_r \nabla_A \delta m_{33} &=D^{(2,1)}\qty(D_A \delta m_{3B} - 2 \lambda' \delta m_{3A})\nonumber\\
    \nabla_r \nabla_A \delta m_{3B} &=D^{(1,2)}\qty( D_A \delta m_{3B} - \lambda' \delta m_{AB} +\lambda' e^{2 (\lambda - \mu)}\Omega_{AB} \delta m_{33})\nonumber\\
    \nabla_r \nabla_A \delta m_{BC} &=D^{(0,3)}\qty(D_A \delta m_{BC} + \lambda' e^{2(\lambda - \mu)}\qty(\Omega_{AB} \delta m_{3C} + \Omega_{AC} \delta m_{3B}))\nonumber\\
    \nabla_A \nabla_r \delta m_{33} &=D_A D^{(2,1)} \delta m_{33} - 2 \lambda'D^{(1,2)} \delta m_{3A}\\
    \nabla_A \nabla_r \delta m_{3B} &=\lambda' e^{2(\lambda- \mu)}\Omega_{AB} D^{(2,1)} \delta m_{33} + D_A D^{(1,2)} \delta m_{3B} - \lambda' D^{(0,3)} \delta m_{AB}\nonumber\\
    \nabla_A \nabla_r \delta m_{BC} &=D_A D^{(0,3)} \delta m_{BC} + \lambda' e^{2(\lambda - \mu)}\qty(\Omega_{AB} D^{(1,2)} \delta m_{3C} + \Omega_{AC} D^{(1,2)} \delta m_{3B})\nonumber\\
    \nabla_A \nabla_B \delta m_{33} &=D_A D_B \delta m_{33} + \lambda' e^{2(\lambda - \mu)} \Omega_{AB} D^{(2,2)} \delta m_{33} - 4 \lambda' D_{(A}\delta m_{B)r} + 2 (\lambda')^2 \delta m_{AB}\nonumber\\
    \nabla_A \nabla_B \delta m_{3C} &= 2 \lambda' e^{2(\lambda - \mu)} \Omega_{C(A}D_{B)} \delta m_{33} + D_A D_B \delta m_{3C} + \lambda' e^{2(\lambda - \mu)} \Omega_{AB} D^{(1,2)} \delta m_{3C}\nonumber\\
    &- 2 (\lambda')^2 e^{2(\lambda - \mu)} \Omega_{AC} \delta m_{3B} - (\lambda')^2 e^{2(\lambda - \mu)} \Omega_{BC} \delta m_{3A} - 2 \lambda' D_{(A} \delta m_{B)C}\nonumber\\
    \nabla_A \nabla_B \delta m_{CD} &=(\lambda')^2 e^{4(\lambda - \mu)}\qty(\Omega_{BC} \Omega_{AD}+ \Omega_{BD} \Omega_{AC}) \delta m_{33}+ 2 \lambda' e^{2(\lambda - \mu)}\qty(\Omega_{C(A}D_{B)}\delta m_{3D} + \Omega_{D(A} D_{B)} \delta m_{3C})\nonumber\\
    &+ D_A D_B \delta m_{CD}+ \lambda' e^{2(\lambda - \mu)} \qty(\Omega_{AB} D^{(0,2)} \delta m_{CD} - 2 \lambda' \Omega_{A(C}\delta m_{D)B})\nonumber
\end{align}

We calculate the splitting of the second order Hamiltonian constraint in four steps. First we consider the momentum terms, i.e. the first three lines of \eqref{eq:Ham2ndOrder}. Their expansion into radial and angular components is
\begin{align}
    &\frac{1}{2} e^{4 \mu} \delta W^{33}\delta W^{33} + 2 e^{2(\mu + \lambda)} \Omega_{AB}\delta W^{3A} \delta W^{3B} + e^{4 \lambda} \qty(\delta W^{AB} \delta W_{AB} - \frac{1}{2} (\delta W^A{}_A)^2) \nonumber\\
    &- \Omega_{AB} e^{2(\mu + \lambda)} \delta W^{33} \delta W^{AB}+ \frac{1}{4}(3 \pi_\mu - \pi_\lambda) \delta W^{33} \delta m_{33}  - \delta W^{33} \Omega^{AB} \delta m_{AB} \frac{\pi_\mu}{4} e^{2(\mu - \lambda)} + \pi_\mu \delta W^{3A} \delta m_{3A}\nonumber\\
    &- \Omega_{AB} \delta W^{AB} \delta m_{33} \frac{\pi_\mu}{4} e^{2(\lambda - \mu)}- \frac{1}{2}\delta W^{AB} \delta m_{AB} (\pi_\mu - \pi_\lambda)+ \frac{1}{4}(\pi_\mu - \pi_\lambda) \Omega_{AB}\Omega^{CD} \delta W^{AB} \delta m_{CD}\\
    &+\qty( \frac{3}{64}\pi_\mu^2 + \frac{1}{32}\pi_\mu \pi_\lambda)e^{-4\mu}(\delta m_{33})^2 + \qty(\frac{1}{16} \pi_\mu^2 + \frac{1}{8}\pi_\mu \pi_\lambda)e^{-2(\mu + \lambda)} \Omega^{AB}\delta m_{3A}\delta m_{3B}\nonumber\\
    &- \frac{3}{32} \pi_\mu^2 e^{-2(\mu + \lambda)} \delta m_{33} \Omega^{AB} \delta m_{AB}+\qty(\frac{\pi_\mu^2}{32} - \frac{\pi_\mu \pi_\lambda}{16} + \frac{\pi_\lambda^2}{16}) e^{-4 \lambda} \Omega^{AB} \Omega^{CD} \delta m_{AC} \delta m_{BD}\nonumber\\
    &+\qty(\frac{\pi_\mu^2}{64}  + \frac{\pi_\mu \pi_\lambda}{32} - \frac{\pi_\lambda^2}{32}) e^{- 4\lambda} \Omega^{AB} \Omega^{CD} \delta m_{AB} \delta m_{CD}\nonumber\,.
\end{align}

The next line in \eqref{eq:Ham2ndOrder} containing the curvature tensors of the background gives
\begin{align}
    &- \frac{1}{4}\qty(1 + \qty((\lambda')^2  + 2\lambda'' - 2 \lambda'\mu'))e^{2 (\lambda - \mu)}) e^{-4 \mu - 2 \lambda} (\delta m_{33})^2\nonumber\\
    &+ \frac{1}{2} \qty(\lambda'' + \lambda'(\lambda' - \mu'))e^{-2 \lambda - 4 \mu}\delta m_{33} \Omega^{AB} \delta m_{AB}-(\lambda'' + \lambda'(\lambda' - \mu')) e^{-2 \lambda - 4 \mu} \Omega^{AB} \delta m_{3A} \delta m_{3B}\\
    &+ \frac{1}{2}\qty(1 - (\lambda')^2e^{2(\lambda - \mu)})e^{-6 \lambda} \delta m^{AB} \delta m_{AB} - \frac{1}{4} \qty(1 - (\lambda')^2e^{2(\lambda - \mu)})e^{- 6\lambda} \qty(\delta m_{AB} \Omega^{AB})^2\nonumber
\end{align}

The following term in \eqref{eq:Ham2ndOrder} are the first order contributions. They are given by
\small
\begin{align}
    &- \sqrt{\overline m}\nabla_\mu \delta m_{\nu \rho} \nabla_\sigma \delta m_{\alpha \beta} \overline m_{(1)}^{\mu \nu \rho \sigma \alpha \beta}= \frac{1}{2}e^{-4 \mu - 2\lambda}(\partial_r - 2 \mu')\delta m_{33}(\partial_r - 2\lambda')\delta m_{AB}\Omega^{AB}\nonumber\\
    &- e^{-4\mu - 2\lambda}(\partial_r - 2 \mu')\delta m_{33}\qty(D^A \delta m_{3A} - \lambda' \Omega^{AB} \delta m_{AB} + 2\lambda' e^{2(\lambda - \mu)} \delta m_{33} )\nonumber\\
    &+\frac{1}{2} e^{-4\mu -2 \lambda} \qty(D_A \delta m_{33} - 2 \lambda' \delta m_{3A})\qty(D_B \delta m_{33} - 2 \lambda' \delta m_{3B})\Omega^{AB}\nonumber\\
    &+e^{-2\mu - 4 \lambda} \qty(D_A \delta m_{BC} + \lambda' e^{2(\lambda - \mu)}\qty(\Omega_{AB} \delta m_{3C} + \Omega_{AC} \delta m_{3B}))\qty(D_D \delta m_{33} - 2 \lambda' \delta m_{3D}) \qty(\Omega^{AB} \Omega^{CD} - \frac{1}{2} \Omega^{AD} \Omega^{BC})\nonumber\\
    &+e^{-2\mu - 4 \lambda}\qty(D_A \delta m_{BC} + \lambda' e^{2(\lambda - \mu)}\qty(\Omega_{AB} \delta m_{3C} + \Omega_{AC} \delta m_{3B}))(\partial_r - \mu' - \lambda')\delta m_{3D}(\Omega^{AD}\Omega^{BC}-2\Omega^{AB}\Omega^{CD})\nonumber\\
    &+e^{-2\mu - 4\lambda}\qty(D_A \delta m_{3B} - \lambda' \delta m_{AB} + \lambda' e^{2(\lambda-\mu)} \Omega_{AB} \delta m_{33})\qty(D_C \delta m_{3D} - \lambda' \delta m_{CD} + \lambda' e^{2(\lambda-\mu)} \Omega_{CD} \delta m_{33})\times\\
    &~~~~~~~~~~~~~~~~~~~~~~~~~~~~~~~~~~~~~~~~~\times \qty(\frac{3}{2} \Omega^{AC} \Omega^{BD} - \frac{1}{2} \Omega^{AD}\Omega^{BC} - \Omega^{AB} \Omega^{CD})\nonumber\\
    &+e^{-2\mu - 4 \lambda}(\partial_r - 2 \lambda')\delta m_{AB}\qty(D_C \delta m_{3D} - \lambda' \delta m_{CD} + \lambda' e^{2(\lambda - \mu)}\Omega_{CD} \delta m_{33})(\Omega^{AB} \Omega^{CD} - \Omega^{AD} \Omega^{BC})\nonumber\\
    &+\frac{1}{4}e^{-2\mu -4 \lambda}(\partial_r -2 \lambda')\delta m_{AB} (\partial_r - 2 \lambda')\delta m_{CD}(3 \Omega^{AC} \Omega^{BD}-\Omega^{AB}\Omega^{CD})\nonumber\\
    &+ e^{-6\lambda} \qty(D_A \delta m_{BC} + \lambda' e^{2(\lambda - \mu)}(\Omega_{AB} \delta m_{3C} + \Omega_{AC}\delta m_{3B} ))\qty(D_D \delta m_{EF} + \lambda' e^{2(\lambda - \mu)}(\Omega_{DE} \delta m_{3F} + \Omega_{DF}\delta m_{3E} ))\overline m^{ABCDEF}_{(1)}\nonumber
\end{align}
\normalsize
The last term consists of second derivatives and is given by
\small
\begin{align}
    &- \sqrt{\overline m} \delta m_{\mu \nu} \nabla_\rho \nabla_\sigma \delta m_{\alpha \beta} \overline m_{(2)}^{\mu \nu \rho \sigma \alpha \beta}\nonumber\\
    &=e^{-4 \mu - 2 \lambda} \Big[\frac{1}{2}\delta m_{33} \nabla_r \nabla_r \delta m_{AB} \Omega^{AB} - \frac{1}{2} \delta m_{33} \nabla_r \nabla_A \delta m_{3B} \Omega^{AB}  - \frac{1}{2} \delta m_{33} \nabla_A \nabla_r \delta m_{3B} \Omega^{AB} +\frac{1}{2}\delta m_{33}\nabla_A \nabla_B \delta m_{33} \Omega^{AB} \Big]\nonumber\\
    &+ e^{-2\mu - 4 \lambda}\Big[ \frac{1}{2}\delta m_{33} \nabla_A \nabla_B \delta m_{CD}(\Omega^{AC}\Omega^{BD} - \Omega^{AB}\Omega^{CD}) + \delta m_{3A} \nabla_r \nabla_B \delta m_{CD}(\Omega^{AB} \Omega^{CD}-\Omega^{AC}\Omega^{BD})\nonumber\\
    &+\delta m_{3A} \nabla_B \nabla_C \delta m_{3D}(2\Omega^{AD}\Omega^{BC} - \Omega^{AC} \Omega^{BD} - \Omega^{AB} \Omega^{CD}) + \delta m_{3A} \nabla_B \nabla_r \delta m_{CD}(\Omega^{AB} \Omega^{CD} - \Omega^{AC} \Omega^{BD}\\
    &+\delta m_{AB} \nabla_r \nabla_r \delta m_{CD}\qty(\Omega^{AC} \Omega^{BD} - \frac{1}{2}\Omega^{AB}\Omega^{CD}) + \delta m_{AB} \nabla_r \nabla_C \delta m_{3D}\qty(\frac{1}{2} \Omega^{AB} \Omega^{CD} - \Omega^{AC} \Omega^{BD})\nonumber\\
    &+ \delta m_{AB} \nabla_C \nabla_r \delta m_{3D}\qty(\frac{1}{2} \Omega^{AB} \Omega^{CD} - \Omega^{AC} \Omega^{BD}) + \delta m_{AB} \nabla_C \nabla_D \delta m_{33}\qty(\Omega^{AC} \Omega^{BD} - \frac{1}{2}\Omega^{AB} \Omega^{CD})\Big]\nonumber\\
    &+e^{-6\lambda} \delta m_{AB} \nabla_C \nabla_D \delta m_{EF} \overline m_{(2)}^{ABCDEF}\nonumber
\end{align}
\normalsize
To derive the formulas of the main text we insert the definition of the perturbed metric and the perturbed momentum. By an integration over the sphere we determine the symmetric constraints shown in \eqref{eq:Ch2} and \eqref{eq:Cv2}.

In the remaining part of this section we investigate the electromagnetic contributions to the constraints. The Gauß constraint $V_G$ is given by
\begin{align}
    V_G = D_\mu E^\mu = \sqrt{\Omega}\partial_r \xi + \partial_r \delta E^3 + D_A \delta E^A\,,
\end{align}
where we used the form $\overline E^3 = \sqrt{\Omega} \xi$ for the background and introduced the perturbations of the electric field as $\delta E^3$ and $\delta E^A$. 

The electromagnetic part of the diffeomorphism constraint is given by
\begin{align}
    V^\mathrm{em}_\mu = F_{\mu\nu} E^\nu - A_\mu V_G\,,
\end{align}
We only focus on the first term because the second is already done. 
For the computation we introduce the magnetic field $B^\mu := \flatfrac{1}{2} \epsilon^{\mu \nu \rho}F_{\nu \rho}$ with the Levi-Civita symbol $\epsilon^{\mu \nu \rho}$. The inversion of this formula gives  $F_{\mu \nu} = \epsilon_{\mu\nu \rho} B ^\rho$. In terms of the perturbed vector potential, the magnetic field is
\begin{align}
    B^3 &= \sum_{l\geq1,m} \sqrt{l(l+1)} \bm X^{o,lm}_M L_{lm}\\
    B^A &= \sum_{l\geq1,m} \qty(\partial_r \bm X^{o,lm}_M L^A_{e,lm} + (\sqrt{l(l+1)}\bm x_M^{lm} - \partial_r \bm X^{e,lm}_M )L^A_{o,lm})
\end{align}
It is convenient to rewrite the diffeomorphism constraint using the magnetic field:
\begin{align}
    V_\mu^{\mathrm{em}} = F_{\mu \nu} E^\nu = \epsilon_{\mu \nu \rho} E^\nu B^\rho
\end{align}
To determine the first order perturbation of this equation we use the fact that the only non-vanishing component of the background electric field is $\overline E^3$. Due to the antisymmetry of the Levi-Civita symbol only the angular components are non-vanishing:
\begin{align}
    V_A^{\mathrm{em}} = - \sqrt{\Omega} \xi \epsilon_{AC} B^C =  \sqrt{\Omega} \xi \sum_{l\geq 1, m}\qty( - \partial_r \bm X^{o,lm}_M L^o_{lm}+ (\sqrt{l(l+1)}\bm x_M^{lm} - \partial_r \bm X^{e,lm}_M)L^e_{lm})
\end{align}
where $\epsilon_{AC} = \epsilon_{3 AC}$ is the Levi-Civita symbol on the sphere. The radial diffeomorphism constraint to second order is given by
\begin{align}
\begin{split}
    {}^{(2)} C_h &= \int_{S^2} \epsilon_{A C} \delta E^A \delta B^C  \dd\Omega\\
    &= \bm Y_o^M \cdot \partial_r \bm X^o_M + \bm Y_e^M \cdot \partial_r \bm X^e_M - \sqrt{l(l+1)} \bm Y_e^M \cdot \bm x_M\,.
\end{split}
\end{align}

The electromagnetic Hamiltonian constraint $V_0^{\mathrm{em}}$ in terms of the magnetic field is
\begin{align}
    V_0^\mathrm{em} = \frac{1}{2}m_{\mu \nu} \qty( \frac{g^2}{\sqrt{m}} E^\mu E^\nu  + \frac{\sqrt{m}}{g^2} B^\mu B^\nu)\,.
\end{align}
To first order we have
\begin{align}
    {}^{(1)} V_0^\mathrm{em} = \frac{g^2}{2 \sqrt{\overline m}}(\delta m_{33} - \frac{1}{2} \overline m_{33} \delta m_{\mu\nu} \overline m^{\mu \nu}) (\overline E^{3})^2 + \frac{g^2}{\sqrt{\overline m}}\overline m_{33} \overline E^3 \delta E^3\,.
\end{align}
Inserting the definitions of the perturbed metric and electric field we obtain
\begin{equation}
    ({}^{(1)}C_v^\mathrm{em})_{lm} =\sqrt{\Omega} \frac{g^2}{2}e^{-\mu - 2 \lambda}\qty(\qty(\frac{1}{2}\bm x^v_{lm} - \bm x^h_{lm} e^{2(\mu -\lambda)}) \xi^2 + 2 e^{2\mu} \xi \bm y^M_{lm})\,.
\end{equation}
The second order corrections are 
\begin{align}
\begin{split}
    V_0^\mathrm{em} &= \frac{g^2}{2 \sqrt{\overline m}}\qty(\frac{1}{8} \overline m_{3 3}((\delta m^\mu{}_\mu)^2 + 2\delta m_{\mu \nu} \delta m^{\mu \nu}) - \frac{1}{2} \delta m_{33}\delta m^\mu{}_\mu) (\overline E^3)^2\\
    &+\frac{g^2}{\sqrt{\overline m}}(\delta m_{3 \mu} \delta E^\mu \overline E^3- \frac{1}{2} \overline m_{3 3} \delta m_{\mu\nu}\overline m^{\mu \nu} \overline E^3 \delta E^3 ) +\frac{g^2}{2\sqrt{\overline m}}\overline m_{\mu \nu} \delta E^\mu \delta E^\nu + \frac{\sqrt{\overline m}}{2 g^2} \overline m_{\mu \nu} \delta B^\mu \delta B^\nu)\,.
\end{split}
\end{align}
Expanding the equation in terms of the electromagnetic and gravitational variables we obtain
\begin{align}
    {}^{(2)} C_v^\mathrm{em} &= \frac{g^2}{2} e^{-\mu- 2 \lambda} \Big[\frac{1}{8} \big(- e^{-4\mu}\bm x^v \cdot \bm x^v - 4 e^{-2\mu -2 \lambda} \bm x^v \cdot \bm x^h + 4 e^{-2\mu -2 \lambda}(\bm x^o \cdot \bm x^o+ \bm x^e\cdot \bm x^e)+ 8 e^{-4\lambda} \bm x^h \cdot \bm x^h\nonumber\\
    &~~~~~~~~~~~~+ 2 e^{-4\lambda} (\bm X^e\cdot \bm X^e + \bm X^o\cdot \bm X^o)\big) e^{2\mu} \xi^2 + \xi (\bm x^v - 2 e^{2\mu -2\lambda} \bm x^h)\cdot \bm y^M\nonumber\\
    &~~~~~~~~~~~~+ 2(\bm x^o \bm Y_o^M + \bm x^e \bm Y_e^M) \xi+ e^{2\mu} \bm y^M\cdot \bm y^M + e^{2\lambda}\bm Y_e^M \cdot \bm Y_e^M + e^{2\lambda}\bm Y_o^M \cdot \bm Y_o^M\Big]\\
    &+ \frac{1}{2g^2}e^{-\mu - 2 \lambda} \Big[e^{2\mu}l(l+1)\bm X^o_M \cdot \bm X^o_M + e^{2\lambda} \partial_r \bm X^o_M \cdot \partial_r \bm X^o_M\nonumber\\
    &~~~~~~~~~~~~+ e^{2\lambda}(\sqrt{l(l+1)}\bm x_m - \partial_r \bm X_e^M)\cdot(\sqrt{l(l+1)}\bm x_m - \partial_r \bm X_e^M)\Big]\nonumber\,.
\end{align}

\section{Some Formulas for the Even Parity Analysis}
\label{sec:EvenParityAnalysis}
In the last canonical transformation of the even parity sector we shifted the momentum variable $P^e$ by a term equal to $A_\mathrm{grav} Q^e$. The function $A_\mathrm{grav}$ is defined as
\scriptsize 
\begin{align}
    A_\mathrm{grav} &= \frac{2}{(l-1) (l+2) e^{6 \mu + 10\lambda}  (\lambda ')^3 M_1^2 (\pi_\mu^{(0)})^3 \Lambda}\Big[-1024 e^{14 \lambda } \Big(4 g^4 \xi ^4-30 e^{\lambda } g^2 r_s \xi ^2+4 e^{4 \lambda }
   \left(l^2+l-2\right)^2+27 e^{3 \lambda } \left(l^2+l-2\right) r_s\nonumber\\
   &~~~~~~~~~~~~~+e^{2 \lambda } \left(54 r_s^2-7 g^2 \left(l^2+l-2\right)
   \xi ^2\right)\Big) \left(\lambda '\right)^9\nonumber\\
   &+1024 e^{14 \lambda } \left(4 g^4 \xi ^4-27 e^{\lambda } g^2 r_s \xi ^2+4 e^{4
   \lambda } \left(l^2+l-2\right)^2+27 e^{3 \lambda } \left(l^2+l-2\right) r_s+e^{2 \lambda } \left(45 r_s^2-8 g^2
   \left(l^2+l-2\right) \xi ^2\right)\right) \mu ' \left(\lambda '\right)^8\nonumber\\
   &+256 e^{10 \lambda } \Big(6 e^{6 \lambda +2 \mu }
   l^6+18 e^{6 \lambda +2 \mu } l^5+38 e^{6 \lambda +2 \mu } l^4-12 e^{5 \lambda +2 \mu } r_s l^4+4 e^{8 \lambda } \mu ''
   l^4+46 e^{6 \lambda +2 \mu } l^3-24 e^{5 \lambda +2 \mu } r_s l^3+8 e^{8 \lambda } \mu '' l^3 \nonumber\\
   &~~~~~~~~~~~~~-132 e^{6 \lambda +2 \mu }
   l^2-15 e^{2 (\lambda +\mu )} g^4 \xi ^4 l^2-100 e^{4 \lambda +2 \mu } g^2 \xi ^2 l^2-240 e^{4 \lambda +2 \mu } r_s^2
   l^2+384 e^{5 \lambda +2 \mu } r_s l^2+120 e^{3 \lambda +2 \mu } g^2 \xi ^2 r_s l^2\nonumber\\
   &~~~~~~~~~~~~~-12 e^{8 \lambda } \mu '' l^2-8 e^{6
   \lambda } g^2 \xi ^2 \mu '' l^2+24 e^{7 \lambda } r_s \mu '' l^2-152 e^{6 \lambda +2 \mu } l-15 e^{2 (\lambda +\mu )} g^4
   \xi ^4 l-100 e^{4 \lambda +2 \mu } g^2 \xi ^2 l-240 e^{4 \lambda +2 \mu } r_s^2 l\nonumber\\
   &~~~~~~~~~~~~~+396 e^{5 \lambda +2 \mu } r_s l+120 e^{3
   \lambda +2 \mu } g^2 \xi ^2 r_s l-16 e^{8 \lambda } \mu '' l-8 e^{6 \lambda } g^2 \xi ^2 \mu '' l+24 e^{7 \lambda } r_s \mu
   '' l+176 e^{6 \lambda +2 \mu }+12 e^{2 \mu } g^6 \xi ^6+86 e^{2 (\lambda +\mu )} g^4 \xi ^4\nonumber\\
   &~~~~~~~~~~~~~-558 e^{3 \lambda +2 \mu }
   r_s^3+200 e^{4 \lambda +2 \mu } g^2 \xi ^2+1128 e^{4 \lambda +2 \mu } r_s^2+465 e^{2 (\lambda +\mu )} g^2 \xi ^2 r_s^2-4
   e^{4 \lambda } \left(-g^2 \xi ^2+e^{2 \lambda } \left(l^2+l-2\right)+3 e^{\lambda } r_s\right){}^2 \left(\mu '\right)^2\nonumber\\
   &~~~~~~~~~~~~~-744
   e^{5 \lambda +2 \mu } r_s-129 e^{\lambda +2 \mu } g^4 \xi ^4 r_s-624 e^{3 \lambda +2 \mu } g^2 \xi ^2 r_s\nonumber\\
   &~~~~~~~~~~~~~-4 e^{4 \lambda }
   \left(4 g^4 \xi ^4-27 e^{\lambda } g^2 r_s \xi ^2+4 e^{4 \lambda } \left(l^2+l-2\right)^2+27 e^{3 \lambda }
   \left(l^2+l-2\right) r_s+e^{2 \lambda } \left(45 r_s^2-8 g^2 \left(l^2+l-2\right) \xi ^2\right)\right) \lambda ''\nonumber\\
   &~~~~~~~~~~~~~+16 e^{8
   \lambda } \mu ''+4 e^{4 \lambda } g^4 \xi ^4 \mu ''+16 e^{6 \lambda } g^2 \xi ^2 \mu ''+36 e^{6 \lambda } r_s^2 \mu ''-48
   e^{7 \lambda } r_s \mu ''-24 e^{5 \lambda } g^2 \xi ^2 r_s \mu ''\Big) \left(\lambda '\right)^7\nonumber\\
   &-512 e^{10 \lambda }
   \left(-g^2 \xi ^2+e^{2 \lambda } \left(l^2+l-2\right)+3 e^{\lambda } r_s\right) \Big(\mu ' \Big(e^{2 \mu } \Big(-6 g^4
   \xi ^4+44 e^{\lambda } g^2 r_s \xi ^2+e^{4 \lambda } \left(l^4+2 l^3+21 l^2+20 l-44\right)\nonumber\\
   &~~~~~~~~~~~~~-4 e^{3 \lambda } \left(5 l^2+5l-31\right) r_s+e^{2 \lambda } \left(g^2 \left(5 l^2+5 l-34\right) \xi ^2-81 r_s^2\right)\Big)-2 e^{4 \lambda }
   \left(-g^2 \xi ^2+e^{2 \lambda } \left(l^2+l-2\right)+3 e^{\lambda } r_s\right) \lambda ''\Big)\nonumber\\
   &~~~~~~~~~~~~~+2 e^{4 \lambda }
   \left(-g^2 \xi ^2+e^{2 \lambda } \left(l^2+l-2\right)+3 e^{\lambda } r_s\right) \lambda ^{(3)}\Big) \left(\lambda
   '\right)^6\nonumber\\
   &+16 e^{6 \lambda } \Big(48 e^{4 \lambda +2 \mu } \left(g^2 \xi ^2+4 e^{2 \lambda }-4 e^{\lambda } r_s\right)
   \left(\mu '\right)^2 \left(-g^2 \xi ^2+e^{2 \lambda } \left(l^2+l-2\right)+3 e^{\lambda } r_s\right){}^2\nonumber\\
   &~~~~~~~~~~~~~+64 e^{8 \lambda }
   \left(\lambda ''\right)^2 \left(-g^2 \xi ^2+e^{2 \lambda } \left(l^2+l-2\right)+3 e^{\lambda } r_s\right){}^2-32 e^{4
   \lambda +2 \mu } \left(g^2 \xi ^2+4 e^{2 \lambda }-4 e^{\lambda } r_s\right) \mu '' \left(-g^2 \xi ^2+e^{2 \lambda }
   \left(l^2+l-2\right)+3 e^{\lambda } r_s\right){}^2\nonumber\\
   &~~~~~~~~~~~~~+e^{4 \mu } \Big[-36 g^8 \xi ^8+540 e^{\lambda } g^6 r_s \xi ^6+e^{2
   \lambda } g^4 \left(g^2 \left(7 l^2+7 l-430\right) \xi ^2-3024 r_s^2\right) \xi ^4-e^{3 \lambda } g^2 r_s \left(g^2
   \left(201 l^2+201 l-4754\right) \xi ^2-7500 r_s^2\right) \xi ^2\nonumber\\
   &~~~~~~~~~~~~~+4 e^{8 \lambda } \left(l^2+l-2\right)^2 \left(l^4+2 l^3-51
   l^2-52 l-156\right)+8 e^{7 \lambda } \left(29 l^6+87 l^5-11 l^4-167 l^3-792 l^2-694 l+1548\right) r_s\nonumber\\
   &~~~~~~~~~~~~~-4 e^{5 \lambda } r_s
   \left(g^2 \left(107 l^4+214 l^3+737 l^2+630 l-3320\right) \xi ^2+18 \left(27 l^2+27 l-298\right) r_s^2\right)\nonumber\\
   &~~~~~~~~~~~~~+4 e^{6
   \lambda } \left(g^2 \left(-15 l^6-45 l^5+37 l^4+149 l^3+394 l^2+312 l-832\right) \xi ^2+24 \left(6 l^4+12 l^3+67 l^2+61
   l-254\right) r_s^2\right)\nonumber\\
   &~~~~~~~~~~~~~-4 e^{4 \lambda } \left(-2 g^4 \left(9 l^4+18 l^3+40 l^2+31 l-226\right) \xi ^4+g^2 \left(-293
   l^2-293 l+4378\right) r_s^2 \xi ^2+1737 r_s^4\right)\Big]\nonumber\\
   &~~~~~~~~~~~~~+32 e^{4 \lambda +2 \mu } \Big[6 g^6 \xi ^6-62 e^{\lambda } g^4
   r_s \xi ^4+e^{6 \lambda } \left(l^2+l-2\right)^2 \left(l^2+l+22\right)-e^{5 \lambda } \left(17 l^4+34 l^3-207 l^2-224
   l+380\right) r_s\nonumber\\
   &~~~~~~~~~~~~~-e^{3 \lambda } r_s \left(g^2 \left(-79 l^2-79 l+314\right) \xi ^2+243 r_s^2\right)+e^{4 \lambda } \left(4
   g^2 \left(l^4+2 l^3-15 l^2-16 l+28\right) \xi ^2-3 \left(47 l^2+47 l-178\right) r_s^2\right)\nonumber\\
   &~~~~~~~~~~~~~+e^{2 \lambda } \left(g^4
   \left(-11 l^2-11 l+46\right) \xi ^4+213 g^2 r_s^2 \xi ^2\right)\Big] \lambda ''\Big) \left(\lambda '\right)^5\\
   &+16 e^{6
   \lambda +2 \mu } \left(g^2 \xi ^2+4 e^{2 \lambda }-4 e^{\lambda } r_s\right) \left(-g^2 \xi ^2+e^{2 \lambda }
   \left(l^2+l-2\right)+3 e^{\lambda } r_s\right) \Big[\mu ' \Big(e^{2 \mu } \Big(-36 g^4 \xi ^4+267 e^{\lambda } g^2 r_s
   \xi ^2\nonumber\\
   &~~~~~~~~~~~~~+16 e^{4 \lambda } \left(l^4+2 l^3+8 l^2+7 l-18\right)-12 e^{3 \lambda } \left(8 l^2+8 l-65\right) r_s-4 e^{2 \lambda
   } \left(g^2 \left(-5 l^2-5 l+54\right) \xi ^2+123 r_s^2\right)\Big)\nonumber\\
   &~~~~~~~~~~~~~-64 e^{4 \lambda } \left(-g^2 \xi ^2+e^{2 \lambda }
   \left(l^2+l-2\right)+3 e^{\lambda } r_s\right) \lambda ''\Big)+32 e^{4 \lambda } \left(-g^2 \xi ^2+e^{2 \lambda }
   \left(l^2+l-2\right)+3 e^{\lambda } r_s\right) \lambda ^{(3)}\Big] \left(\lambda '\right)^4\nonumber\\
   &-8 e^{2 (\lambda +\mu )}
   \left(g^2 \xi ^2+4 e^{2 \lambda }-4 e^{\lambda } r_s\right) \Big[18 e^{4 \lambda +2 \mu } \left(g^2 \xi ^2+4 e^{2 \lambda
   }-4 e^{\lambda } r_s\right) \left(\mu '\right)^2 \left(-g^2 \xi ^2+e^{2 \lambda } \left(l^2+l-2\right)+3 e^{\lambda }
   r_s\right){}^2\nonumber\\
   &~~~~~~~~~~~~~+32 e^{8 \lambda } \left(\lambda ''\right)^2 \left(-g^2 \xi ^2+e^{2 \lambda } \left(l^2+l-2\right)+3
   e^{\lambda } r_s\right){}^2-10 e^{4 \lambda +2 \mu } \left(g^2 \xi ^2+4 e^{2 \lambda }-4 e^{\lambda } r_s\right) \mu ''
   \left(-g^2 \xi ^2+e^{2 \lambda } \left(l^2+l-2\right)+3 e^{\lambda } r_s\right){}^2\nonumber\\
   &~~~~~~~~~~~~~+e^{4 \mu } \Big[-4 g^8 \xi ^8+64
   e^{\lambda } g^6 r_s \xi ^6-e^{2 \lambda } g^4 \left(g^2 \left(11 l^2+11 l+54\right) \xi ^2+374 r_s^2\right) \xi ^4+2 e^{3
   \lambda } g^2 r_s \left(5 g^2 \left(8 l^2+8 l+61\right) \xi ^2+477 r_s^2\right) \xi ^2\nonumber\\
   &~~~~~~~~~~~~~+4 e^{8 \lambda }
   \left(l^2+l-2\right)^2 \left(l^4+2 l^3-15 l^2-16 l-24\right)+4 e^{7 \lambda } \left(22 l^6+66 l^5-49 l^4-208 l^3-189 l^2-74
   l+432\right) r_s\nonumber\\
   &~~~~~~~~~~~~~+e^{5 \lambda } r_s \left(g^2 \left(-199 l^4-398 l^3-73 l^2+126 l+1768\right) \xi ^2+36
   \left(l^2+l+78\right) r_s^2\right)\nonumber\\
   &~~~~~~~~~~~~~+4 e^{6 \lambda } \left(g^2 \left(-6 l^6-18 l^5+23 l^4+76 l^3+41 l^2-116\right) \xi
   ^2+\left(73 l^4+146 l^3+108 l^2+35 l-812\right) r_s^2\right)\nonumber\\
   &~~~~~~~~~~~~~-e^{4 \lambda } \left(16 g^4 \left(-2 l^4-4 l^3+l^2+3
   l+15\right) \xi ^4+g^2 \left(155 l^2+155 l+2276\right) r_s^2 \xi ^2+900 r_s^4\right)\Big]\nonumber\\
   &~~~~~~~~~~~~~+2 e^{4 \lambda +2 \mu }
   \Big[36 g^6 \xi ^6-375 e^{\lambda } g^4 r_s \xi ^4+16 e^{6 \lambda } \left(l^2+l-2\right)^2 \left(l^2+l+9\right)-12 e^{5
   \lambda } \left(4 l^4+8 l^3-105 l^2-109 l+202\right) r_s\nonumber\\
   &~~~~~~~~~~~~~-9 e^{3 \lambda } r_s \left(g^2 \left(-47 l^2-47 l+218\right) \xi
   ^2+164 r_s^2\right)-4 e^{4 \lambda } \left(3 \left(65 l^2+65 l-277\right) r_s^2-g^2 \left(l^4+2 l^3-91 l^2-92 l+180\right)
   \xi ^2\right)\nonumber\\
   &~~~~~~~~~~~~~+e^{2 \lambda } \left(1293 g^2 \xi ^2 r_s^2-8 g^4 \left(7 l^2+7 l-36\right) \xi ^4\right)\Big] \lambda
   ''\Big] \left(\lambda '\right)^3\nonumber\\
   &-4 e^{2 \lambda +4 \mu } \left(g^2 \xi ^2+4 e^{2 \lambda }-4 e^{\lambda } r_s\right){}^2
   \left(-g^2 \xi ^2+e^{2 \lambda } \left(l^2+l-2\right)+3 e^{\lambda } r_s\right) \Big[\mu ' \Big(e^{2 \mu } \Big(-8 g^4
   \xi ^4+63 e^{\lambda } g^2 r_s \xi ^2\nonumber\\
   &~~~~~~~~~~~~~+8 e^{4 \lambda } \left(l^4+2 l^3+4 l^2+3 l-10\right)-12 e^{3 \lambda }
   \left(l^2+l-17\right) r_s-8 e^{2 \lambda } \left(7 g^2 \xi ^2+15 r_s^2\right)\Big)\nonumber\\
   &~~~~~~~~~~~~~-52 e^{4 \lambda } \left(-g^2 \xi
   ^2+e^{2 \lambda } \left(l^2+l-2\right)+3 e^{\lambda } r_s\right) \lambda ''\Big)+20 e^{4 \lambda } \left(-g^2 \xi ^2+e^{2
   \lambda } \left(l^2+l-2\right)+3 e^{\lambda } r_s\right) \lambda ^{(3)}\Big] \left(\lambda '\right)^2\nonumber\\
   &+e^{4 \mu }
   \left(g^2 \xi ^2+4 e^{2 \lambda }-4 e^{\lambda } r_s\right){}^2 \Big[8 e^{2 (\lambda +\mu )} \left(g^2 \xi ^2+4 e^{2
   \lambda }-4 e^{\lambda } r_s\right) \left(\mu '\right)^2 \left(-g^2 \xi ^2+e^{2 \lambda } \left(l^2+l-2\right)+3 e^{\lambda
   } r_s\right){}^2\nonumber\\
   &~~~~~~~~~~~~~+16 e^{6 \lambda } \left(\lambda ''\right)^2 \left(-g^2 \xi ^2+e^{2 \lambda } \left(l^2+l-2\right)+3
   e^{\lambda } r_s\right){}^2\nonumber\\
   &~~~~~~~~~~~~~-4 e^{2 (\lambda +\mu )} \left(g^2 \xi ^2+4 e^{2 \lambda }-4 e^{\lambda } r_s\right) \mu ''
   \left(-g^2 \xi ^2+e^{2 \lambda } \left(l^2+l-2\right)+3 e^{\lambda } r_s\right){}^2\nonumber\\
   &~~~~~~~~~~~~~+e^{4 \mu } \Big[g^6 \left(-5 l^2-5
   l+2\right) \xi ^6+e^{\lambda } g^4 \left(45 l^2+45 l-22\right) r_s \xi ^4+2 e^{2 \lambda } g^2 \Big(2 g^2 \left(3 l^4+6
   l^3-7 l^2-10 l+4\right) \xi ^2-5 \left(13 l^2+13 l-8\right) r_s^2\Big) \xi ^2\nonumber\\
   &~~~~~~~~~~~~~+4 e^{6 \lambda } l \left(l^2+l-2\right)^2
   \left(l^3+2 l^2-3 l-4\right)+8 e^{5 \lambda } \left(l^2+l-2\right)^2 \left(5 l^2+5 l-3\right) r_s\nonumber\\
   &~~~~~~~~~~~~~+2 e^{3 \lambda } r_s
   \left(g^2 \left(-37 l^4-74 l^3+77 l^2+114 l-56\right) \xi ^2+12 \left(5 l^2+5 l-4\right) r_s^2\right)\nonumber\\
   &~~~~~~~~~~~~~+4 e^{4 \lambda }
   \left(l^2+l-2\right) \left(g^2 \left(-3 l^4-6 l^3+7 l^2+10 l-4\right) \xi ^2+3 \left(9 l^2+9 l-8\right)
   r_s^2\right)\Big]\nonumber\\
   &~~~~~~~~~~~~~+4 e^{2 (\lambda +\mu )} \Big[8 g^6 \xi ^6-87 e^{\lambda } g^4 r_s \xi ^4+8 e^{6 \lambda }
   \left(l^2+l-2\right)^2 \left(l^2+l+5\right)+12 e^{5 \lambda } \left(l^4+2 l^3+26 l^2+25 l-54\right) r_s\nonumber\\
   &~~~~~~~~~~~~~-3 e^{3 \lambda }
   r_s \left(g^2 \left(-25 l^2-25 l+166\right) \xi ^2+120 r_s^2\right)-4 e^{4 \lambda } \left(2 g^2 \left(l^4+2 l^3+11 l^2+10
   l-24\right) \xi ^2+3 \left(13 l^2+13 l-71\right) r_s^2\right)\nonumber\\
   &~~~~~~~~~~~~~+e^{2 \lambda } \left(309 g^2 \xi ^2 r_s^2-8 g^4
   \left(l^2+l-9\right) \xi ^4\right)\Big] \lambda ''\Big] \lambda '\nonumber\\
   &-4 e^{2 \lambda +6 \mu } \left(g^2 \xi ^2+4 e^{2
   \lambda }-4 e^{\lambda } r_s\right){}^3 \left(-g^2 \xi ^2+e^{2 \lambda } \left(l^2+l-2\right)+3 e^{\lambda } r_s\right){}^2
   \left(3 \mu ' \lambda ''-\lambda ^{(3)}\right)\Big]\nonumber
\end{align}
\normalsize
The boundary term which arises by simplifying the solution for $\pi_\mu^{(2)}$ is
\scriptsize
\begin{align}
    &\qty(2 e^{-\lambda -2 \mu }-\frac{3 e^{-3\lambda}}{\lambda'^2} \Delta) (q_2')^2 + \frac{e^{-4 \lambda } \left(r_s-e^{\lambda } \left(l^2+l+2\right)\right)}{\lambda '} q_2' q_2 + \qty(4 e^{\lambda -4 \mu } \lambda '- \frac{2 e^{-2\mu -\lambda}}{\lambda'}\Delta) q_2' q_1 - \frac{3 }{4 \lambda'} e^{-3\lambda} \pi_\mu^{(0)} q_2' p_1 -\frac{1}{2} e^{-3 \lambda } \sqrt{l (l+1)} g^2 \xi A q_2\nonumber\\
    &+ \frac{4 e^{2 \mu -2 \lambda } \left(e^{\lambda } \left(l^2+l+2\right)-r_s\right) \left(-20 e^{4 \lambda +2 \mu } \left(\lambda '\right)^2 \left(g^2
   \xi ^2+4 e^{2 \lambda }-4 e^{\lambda } r_s\right)+e^{4 \mu } \left(g^2 \xi ^2+4 e^{2 \lambda }-4 e^{\lambda } r_s\right){}^2+64 e^{8 \lambda }
   \left(\lambda '\right)^4\right)}{\pi _0(r) \left(e^{2 \mu } \left(g^2 \xi ^2+4 e^{2 \lambda }-4 e^{\lambda } r_s\right)-8 e^{4 \lambda }
   \left(\lambda '\right)^2\right){}^2} q_2 p_1 \nonumber\\
   &+\Bigg[\frac{4 e^{-3\mu - 3 \lambda} \left(e^{\lambda } \left(l^2+l+2\right)-r_s\right) \Delta }{\lambda' M_1}+\frac{1}{4} e^{-3 \lambda -2
   \mu } \left(g^2 \xi ^2-2 e^{2 \lambda } l (l+1)-2 e^{\lambda } r_s\right)-\frac{4 e^{- \lambda -2 \mu } \left(\lambda ' \mu '-\lambda ''\right) \Delta}{4 \left(\lambda '\right)^2}-2 e^{\lambda -4 \mu } \left(\lambda ''-\lambda '
   \mu '+\left(\lambda '\right)^2\right)\Bigg] q_2 q_1\nonumber\\
   & + \Bigg[-\frac{4 l (l+1) e^{- 7\lambda } \Delta^2 \Lambda}{(\lambda')^2 M_1^2}+\frac{1}{2}e^{-4 \lambda } \left(4 r_s-3 e^{\lambda }
   \left(l^2+l+2\right)\right)+\frac{e^{-10\lambda - 6\mu}}{4 (\lambda')^2 M_1^2} \Big(16 e^{4 \lambda +2 \mu } \left(\lambda '\right)^2 \Big(-2
   e^{\lambda } \left(6 g^2 \xi ^2+17 r_s^2\right)+9 g^2 \xi ^2 r_s\nonumber\\
   &~~~~~~~~~+4 e^{3 \lambda } \left(l^2+l-10\right)-2 e^{2 \lambda } \left(l^2+l-38\right)
   r_s\Big)+e^{4 \mu } \left(g^2 \xi ^2+4 e^{\lambda } \left(e^{\lambda }-r_s\right)\right) \Big(6 e^{\lambda } \left(6 r_s^2-g^2 \left(l^2+l-2\right) \xi ^2\right)-9 g^2 \xi ^2 r_s\nonumber\\
   &~~~~~~~~~+8 e^{3 \lambda } \left(l^4+2 l^3-l+6\right)+4 e^{2 \lambda } (4 l (l+1)-21) r_s\Big)+64 e^{8
   \lambda } \left(\lambda '\right)^4 \left(2 e^{\lambda } \left(l^2+l+6\right)-9 r_s\right)\Big]\Bigg] (q_2)^2\nonumber\\
   &+\Bigg[\frac{g^2 \xi  e^{-3\lambda -6 \mu }}{8 \sqrt{l
   (l+1)}\Lambda^2}\Big[e^{2 \mu } \lambda '' \left(g^2 \xi ^2+4 e^{\lambda } \left(e^{\lambda }-r_s\right)\right) \left(-g^2
   \xi ^2+e^{2 \lambda } \left(l^2+l-2\right)+3 e^{\lambda } r_s\right)+8 e^{4 \lambda } \left(\lambda '\right)^3 \mu ' \left(-g^2 \xi ^2+e^{2 \lambda
   } \left(l^2+l-2\right)+3 e^{\lambda } r_s\right)\nonumber\\
   &~~~~~~~~~-e^{2 \mu } \lambda ' \mu ' \left(g^2 \xi ^2+4 e^{\lambda } \left(e^{\lambda }-r_s\right)\right)
   \left(-g^2 \xi ^2+e^{2 \lambda } \left(l^2+l-2\right)+3 e^{\lambda } r_s\right)-8 e^{4 \lambda } \left(\lambda '\right)^4 \left(-3 g^2 \xi ^2+e^{2
   \lambda } \left(l^2+l-2\right)+6 e^{\lambda } r_s\right)\nonumber\\
   &~~~~~~~~~+\left(\lambda '\right)^2 \Big(e^{2 \mu } \left(-3 g^4 \xi ^4-e^{2 \lambda } \left(g^2 \left(l^2+l+18\right) \xi ^2+30
   r_s^2\right)+20 g^2 e^{\lambda } \xi ^2 r_s+2 e^{4 \lambda } (l-1) (l+2) \left(l^2+l+4\right)+48 e^{3 \lambda } r_s\right)\nonumber\\
   &~~~~~~~~~+ 8 e^{4 \lambda } \lambda '' \left(g^2 \xi ^2-e^{2 \lambda }
   \left(l^2+l-2\right)-3 e^{\lambda } r_s\right)\Big)\Bigg] A q_1\\
   &+ \frac{e^{-4 \mu + \lambda } \left(-192 e^{2 \mu - 2 \lambda} \left(\lambda '\right)^2  \Delta + 48 e^{4 \mu - 4 \lambda}  \Delta^2+128 \left(\lambda
   '\right)^4\right)}{2(\lambda')^2 M_1^2}(p_1)^2 + \frac{2e^{-5 \mu }\pi_\mu^{(0)} \left(4 \left(\lambda '\right)^2-e^{2 \mu - 2\lambda} \Delta\right)}{\lambda' M_1} q_1 p_1 \nonumber\\
   &+ \frac{g^2 e^{-2 \mu } \xi}{4 \sqrt{l^2+l-2}}  \Bigg[-\frac{2 g^2 \xi^2 e^{-2\lambda} \left(\lambda '\right)^2}{\Lambda}+\frac{e^{-4 \lambda }}{2 \Lambda \lambda' e^{3\mu + 3 \lambda} M_1} \Big(-32 e^{6 \lambda } \left(\lambda '\right)^4 \left(-2 g^2 \xi ^2+e^{2 \lambda }
   \left(l^2+l-2\right)+6 e^{\lambda } r_s\right)\nonumber\\
   &~~~~~~~~~+e^{4 \mu } \left(g^2 \xi ^2+4 e^{\lambda } \left(e^{\lambda }-r_s\right)\right) \left(-g^2
   \left(l^2+l+2\right) \xi ^2+2 e^{2 \lambda } (l-1) (l+2) \left(l^2+l+2\right)+2 e^{\lambda } \left(l^2+l+4\right) r_s\right)\nonumber\\
   &~~~~~~~~~-4 e^{2 (\lambda +\mu
   )} \left(\lambda '\right)^2 \left(2 g^4 \xi ^4+e^{2 \lambda } \left(30 r_s^2-g^2 \left(l^2+l-10\right) \xi ^2\right)-16 g^2 e^{\lambda } \xi ^2
   r_s+4 e^{3 \lambda } \left(l^2+l-8\right) r_s+2 e^{4 \lambda } (l-1) l (l+1) (l+2)\right)\Big)\nonumber\\
   &~~~~~~~~~+4 \lambda ''-4 \lambda ' \mu '\Bigg] Q^e A^e \nonumber + \frac{e^{-2 \lambda } g^4 \xi^2 \Delta }{4 \Lambda} (A^e)^2 + \qty(\frac{2(l^2+l+2)r }{l(l+1)(l+2)(l-1)} + O (r^0)) (q_1)^2 + \qty(-\frac{1}{r} + O(r^{-2})) (Q^e)^2
\end{align}
\normalsize
The expressions for the terms in front of $(q_1)^2$ and $(Q^e)^2$ in the last line are very long and would require too much space if shown explicitly.  

\normalsize

\bibliography{references}

\providecommand{\href}[2]{#2}\begingroup\raggedright\begin{thebibliography}{10}

\bibitem{II}
J.~Neuser and T.~Thiemann, ``Quantum field theory of black hole perturbations
  with backreaction spherically symmetric 2nd order einstein sector,''.

\bibitem{III}
J.~Neuser and T.~Thiemann, ``Quantum field theory of black hole perturbations
  with backreaction spherically symmetric 2nd order maxwell sector,''.

\bibitem{I}
T.~Thiemann, ``Quantum field theory of black hole perturbations with
  backreaction i. general framework,''.

\bibitem{Chandrasekhar1983}
S.~Chandrasekhar, {\em {The Mathematical Theory of Black Holes}}.
\newblock Oxford University Press, 1983.

\bibitem{6}
V.~Moncrief, ``{Gravitational perturbations of spherically symmetric systems.
  I. The exterior problem},''
  \href{https://dx.doi.org/10.1016/0003-4916(74)90173-0}{{\em Annals of
  Physics} {\bfseries 88} no.~2, (1974) 323--342}.

\bibitem{Moncrief1974a}
V.~Moncrief, ``{Odd-parity stability of a Reissner-Nordstr{\"{o}}m black
  hole},'' \href{https://dx.doi.org/10.1103/PhysRevD.9.2707}{{\em Physical
  Review D} {\bfseries 9} no.~10, (1974) 2707--2709}.

\bibitem{Moncrief1974RN}
V.~Moncrief, ``{Stability of Reissner-Nordstr{\"{o}}m black holes},''
  \href{https://dx.doi.org/10.1103/PhysRevD.10.1057}{{\em Physical Review D}
  {\bfseries 10} no.~4, (Aug, 1974) 1057--1059}.
  \url{https://link.aps.org/doi/10.1103/PhysRevD.10.1057}.

\bibitem{Moncrief1975}
V.~Moncrief, ``{Gauge-invariant perturbations of Reissner-Nordstr{\"{o}}m black
  holes},'' \href{https://dx.doi.org/10.1103/PhysRevD.12.1526}{{\em Physical
  Review D} {\bfseries 12} no.~6, (Sep, 1975) 1526--1537}.
  \url{https://link.aps.org/doi/10.1103/PhysRevD.12.1526}.

\bibitem{7}
D.~Brizuela and J.~M. Mart{\'{i}}n-Garc{\'{i}}a, ``{Hamiltonian theory for the
  axial perturbations of a dynamical spherical background},''
  \href{https://dx.doi.org/10.1088/0264-9381/26/1/015003}{{\em Classical and
  Quantum Gravity} {\bfseries 26} no.~1, (2009) },
  \href{https://arxiv.org/abs/0810.4786}{{\ttfamily arXiv:0810.4786}}.

\bibitem{8}
D.~Brizuela, ``{A generalization of the Zerilli master variable for a dynamical
  spherical spacetime},''
  \href{https://dx.doi.org/10.1088/0264-9381/32/13/135005}{{\em Classical and
  Quantum Gravity} {\bfseries 32} no.~13, (2015) },
  \href{https://arxiv.org/abs/1505.05278}{{\ttfamily arXiv:1505.05278}}.

\bibitem{xAct}
J.~M. Martín-García {\em et~al.}, ``{xAct}: A mathematica package for tensor
  algebra and tensor calculus,'' \url{https://www.xact.es/}.

\bibitem{TT}
T.~Thiemann, ``Symmetry reduction, gauge reduction, backreaction and consistent
  higher order perturbation theory,''.

\end{thebibliography}\endgroup
\bibliographystyle{utphys}

\end{document}